\documentclass[10pt,aps,prb,showpacs,superscriptaddress,twocolumn,
floatfix]{revtex4}
\usepackage{amsmath}
\usepackage{citesort}
\usepackage{graphicx}
\begin{document}
\title{Inversion techniques for optical conductivity data}
\author{E. Schachinger}
\email{schachinger@itp.tu-graz.ac.at}
\homepage{www.itp.tu-graz.ac.at/~ewald}
\affiliation{Institute of Theoretical and Computational Physics\\
Graz University of Technology, A-8010 Graz, Austria}
\author{D. Neuber}
\affiliation{Institute of Theoretical and Computational Physics\\
Graz University of Technology, A-8010 Graz, Austria}
\author{J.P. Carbotte}
\affiliation{Department of Physics and Astronomy, McMaster University,\\
Hamilton, Ontario, L8S 4M1 Canada}
\date{\today}
\begin{abstract}
Optical data is encoded with information on the microscopic
interaction between charge carriers. For an electron-phonon
system, the Eliashberg equations apply and a Kubo formula
can be used to get the infrared conductivity. The task of
extracting the electron-phonon spectral density $\alpha^2F(\omega)$
from data is rather complicated and, thus, simplified but
approximate expressions for the conductivity have often
been used. We test the accuracy of such simplifications and also
discuss the advantages and disadvantages of various numerical
methods needed in the inversion process. Normal and superconducting
state are considered as well as boson exchange mechanisms which
might be applicable to the High-$T_c$ oxides.
\end{abstract}
\maketitle
\section{Introduction}

The interaction Hamiltonian between electrons and phonons involves
a complicated matrix element or coupling function
$g_{{\bf k},{\bf k}',\nu}$ which describes the scattering of an
electron initially in the state $\vert{\bf k}\rangle$ to any final
state $\vert{\bf k}'\rangle$ through the exchange of a phonon
$\omega_\nu({\bf k}'-{\bf k})$. Here $\nu$ is a phonon branch
index and the momentum transfer ${\bf k}'-{\bf k}$ can fall
outside the first Brillouin zone and so phonon Umklapp processes
enter. In a real metal the Bloch states of the band structure
can be complicated and this is reflected in the electronic
state labeled by $\vert{\bf k}\rangle$. Fortunately, many important
properties of an electron-phonon system require for their
understanding only a Fermi surface to Fermi surface average of
the coupling, namely the function\cite{mars3}
\[
  \alpha^2F(\omega) = \frac{1}{N(\mu)}\sum\limits_{{\bf k},{\bf k}'}
  B_\nu({\bf k}'-{\bf k})\delta\left(\varepsilon_{\bf k}-\mu\right)
  \delta\left(\varepsilon_{{\bf k}'}-\mu\right),
\]
where $\mu$ is the chemical potential, $\varepsilon_{\bf k}$
the electron energy, $N(\mu)$ the electron density of states
and $B_\nu({\bf k}'-{\bf k})$ the phonon spectral function.
For example, in the Eliashberg formulation\cite{carb1} of
superconductivity based on Migdal's theorem for electron-phonon
vertex corrections, it is $\alpha^2F(\omega)$ that enters. For
the infrared conductivity another, somewhat different weighting
of $g_{{\bf k},{\bf k}',\nu}$ comes in and the resulting function
of $\omega$ is usually called the transport spectral density
denoted $\alpha^2_{tr}F(\omega)$.\cite{allen,carb4} Here we will
not deal directly with these differences. An important goal of
experiments in conventional superconductors has been to determine
the electron-phonon spectral function $\alpha^2F(\omega)$.\cite{carb1,%
allen} This has been successfully accomplished for a large number
of conventional materials using tunneling data and the inversion
technique of McMillan and Rowell.\cite{mcmillan} In a few cases the
infrared optical conductivity\cite{farnworth,tomlinson} was also
used and excellent agreement with tunneling results was found.

Extensions to the consideration of the A15 compounds revealed that
additional features of the band structure such as the energy dependence
of the electronic density of states $N(\varepsilon)$ can also be
important.\cite{boza1,boza2} More recently the optical data in the
alkali doped C$_{60}$ compounds has been inverted\cite{mars1} and
found consistent with its superconductivity. When experimentally
determined electron-phonon spectral functions are compared with
first principle band structure calculations extended to include
electron-phonon interaction good agreement is obtained.\cite{tomlinson,%
carb1}

In dealing with the high-$T_c$ oxides several complications immediately
arise. First, their superconductivity is not generally believed to be
due to the electron-phonon interaction. A consensus exists that the
gap has $d$-wave rather than $s$-wave symmetry and comes out as a
result of strong correlation effects. A natural explanation for
the $d$-wave gap is found in the antiferromagnetic interaction
certainly present in the cuprates. A possible model is the
Nearly Antiferromagnetic Fermi Liquid model (NAFFL) of Pines and
coworkers.\cite{pines1,pines2} It needs to be pointed out, however,
that there is no a priory reason why the electron-phonon interaction
itself could not lead to a $d$-wave gap and there exists recent work
on this possibility.\cite{kulic1,zeyher,kulic2,weger,weger1,kulic3,%
kulic4} In any case, when
$d$-wave symmetry is involved, the spectral function acting in the
gap channel Eq.~\eqref{eq:B1} need not be the same as the one
that determines the
renormalizations in the $\omega$-channel Eq.~\eqref{eq:B2}. At $T_c$ in
the normal state it is only the latter spectral density that enters.
There exists considerable literature on extensions of Eliashberg theory
to include a $d$-wave gap based on model spectral densities for
the electron-boson interaction that may be involved.\cite{schach5}
Of course, there is no guarantee that the final theory of strongly
correlated systems that is needed to describe the oxides will fall
within the class of boson exchange models. Nevertheless, such
an approach has proven valuable in providing insight into the
physics of the oxides as we will see also in this paper.

In the recent literature, tunneling spectroscopy\cite{zas1,zas2,zas3}
as well as angular resolved photo emission\cite{kam1,campu1,norm,%
eschrig,cuk} has been used to analyze data in terms of boson structure.
Here we wish to concentrate on optical data.\cite{carb3,tu,schach3,hwang}
Optimally doped YBa$_2$Cu$_3$O$_{6.95}$ (YBCO$_{6.95}$) was first considered
within a complete Eliashberg formalism generalized to include
$d$-wave pairing by Carbotte {\it et al.}\cite{carb3} (CSB). A model
form for the electron-boson spectral function coming possibly from
exchange of spin fluctuations and denoted by $I^2\chi(\omega)$ is
assumed with two fitting parameters, the coupling $I^2$ and the
spin fluctuation energy $\omega_{sf}$ in the model of
Millis {\it et al.}\cite{mmp} (MMP) which are varied to get
the best fit to the normal state infrared data at
or close to $T=T_c$. For the
superconducting state the same form of $I^2\chi(\omega)$ is assumed
to also determine the gap channel but its magnitude is different
and is fit to get the measured value of $T_c$. In addition, it is
found that the data in the superconducting state indicates the formation
of an optical resonance in $I^2\chi(\omega)$ not present at $T_c$
which increases in amplitude as $T$ is reduced and is positioned at
$41\,$meV. Similar optical resonances were later found in the
superconducting state of other cuprates although not in all.\cite{schach2}
In some the resonance seems to persist even in the normal state.%
\cite{schach1,schach3} While the work described above involves a
least squares fit of an assumed form for $I^2\chi(\omega)$ to the
optical scattering rate data other inversion techniques\cite{dorde}
have been considered but so far these are based on approximate
analytic formulas for the relationship between the optical
scattering rate and the electron-boson spectral density rather than the
full Eliashberg formulation of Carbotte and coworkers.\cite{mars3,carb3,%
schach5,tu}

Such approximate formulas were given by Allen\cite{allen} for an
electron-phonon system and are based on ordinary second order
perturbation theory at zero temperature. Allen considered the normal
as well as the superconducting state with $s$-wave symmetry. A
generalization to finite temperature was provided by
Shulga {\it et al.}\cite{shulga1} who only considered the normal
state but started directly from an Eliashberg formalism and the
Kubo formula for the conductivity. A generalization to include as well
a pseudogap was recently provided by Sharapov and Carbotte.%
\cite{sharap} Finally, Carbotte and Schachinger\cite{carb2} generalized
the original work of Allen to a superconductor with $d$-wave gap
symmetry.

The advantage of these simplified but approximate equations is that they
relate directly through an integral the optical scattering rate to the
desired spectral function $I^2\chi(\omega)$ and various numerical techniques
such as singular value decomposition\cite{dorde,carb2} can be used
to numerically invert the equation. Of course, the alternate method of
assuming some characteristic functional form for $I^2\chi(\omega)$ and
least squares fit a few parameters to the data can also be employed
based on the simplified equations described above instead of employing
the full Eliashberg equations. For instance, the new equations of
Sharapov and Carbotte\cite{sharap} have already been used in this way
by Hwang {\it et al.}\cite{hwang1} to analyze data in underdoped
Ortho-II YBCO$_{6.5}$.

The aim of this paper is to understand better how limitations in
accuracy of the simplified formulas can impact on the resulting
form of $I^2\chi(\omega)$ and to explore as well the advantages
and limitations of numerical inversion techniques such as SVD
and Maximum Entropy Method (MaxEnt) as well as least squares
fit.

The paper is organized as follows. In Sec.~\ref{sec:2} the formal
background is discussed. One subsection concentrates on the three
major methods of inversion, namely the second derivative method,
deconvolution methods based on approximate relations, and the
least squares fit method. The second subsection discusses approximate
formulas for the normal and superconducting state which allow one to
calculate the optical scattering rate $\tau^{-1}_{op}(\omega)$ from
a given spectral density $\alpha^2F(\omega)$ using a convolution
integral. Section~\ref{sec:3} discusses numerically the caveats and
merits of the various methods of inversion by studying normal
metals as well as High-$T_c$ cuprates. Computer generated and
experimental $\tau^{-1}_{op}(\omega)$ data for the normal and
superconducting state are used as input for the inversion.
Finally, conclusions are drawn in Sec.~\ref{sec:4}. Two
appendices have been added. Appendix~\ref{app:A} gives an
overview of the Maximum Entropy method in terms of Bayesian probability
theory. Appendix~\ref{app:B} presents all important equations
which allow to calculate the optical scattering rate within the
framework of full Eliashberg theory.

\section{Formalism}
\label{sec:2}
\subsection{Methods of Inversion}
\label{ssec:2.1}

In order to understand the mechanism of superconductivity it is
important to have detailed knowledge of the spectral function
$\alpha^2F(\omega)$ and as tunneling, the established source of
information on $\alpha^2F(\omega)$,\cite{mcmillan} was initially not a
successful tool in the high-$T_c$ superconductors, the infrared
optical conductivity, $\sigma_{op}(\omega)$, became increasingly
important, particularly
in the form of the optical scattering rate
\begin{equation}
  \label{eq:1}
  \tau^{-1}_{op}(\omega) = \frac{\Omega_p^2}{4\pi}\Re{\rm e}
  \left[\sigma^{-1}_{op}(\omega)\right]
\end{equation}
of extended Drude theory. Here, $\Omega_p$ is the plasma frequency.

There are, in principle, three methods to extract the information
on $\alpha^2F(\omega)$ from the optical scattering rate (inversion).
An essential requirement for a solution obtained with any
of these methods is
that the result should match the data points as well as
possible. In order to assess
the quality of the fit we need to know how the experimental data points 
$t_i \equiv \tau_{ex}^{-1}(\omega_i)$ scatter around the `true' values 
$t_i^0 \equiv \tau^{-1}_{op}(\omega_i)$, that is, we need to know
in terms of Bayesian probability theory the likelihood
$p({\bf t}\vert{\bf t}^0,{\cal I})$. It describes the distribution 
of $N$ data points ${\bf t} = \{t_i\vert i=1,\ldots,N\}$ given the `exact'
values ${\bf t}^0 = \{t^0_i\vert i=1,\ldots,N\}$, which are usually
expressed in terms of the parameters of the physical model.
The symbol $\mathcal{I}$ 
designates all additionally available background information comprising the
experimental setup as well as the physical model employed. 

The likelihood is determined by the experimental setup and we have to keep in
mind that the experimental signal contains at least three
contributions\cite{sivia}
\begin{equation}
  \label{eq:3}
  \tau^{-1}_{ex}(\omega) = \tau^{-1}_{op}(\omega)+B(\omega)
  \pm \eta(\omega),
\end{equation}
with $B(\omega)$ usually a slowly varying background signal
which is typical of the experimental setup and $\eta(\omega)$, the
noise in the data.

Unfortunately, we do not have any knowledge concerning the functional form of
the likelihood for the experimental data sets considered in this
study. Therefore, we make the \emph{assumption} of an uncorrelated normal
distribution with standard deviation $\sigma$:
\begin{equation}
\label{eq:gaussian_likelihood}
  p({\bf t}\vert{\bf t}^0,{\cal I}) = (2\pi\sigma^2)^{-N/2} 
  \exp\left( -\frac{1}{2}\gamma^2 \right).
\end{equation}
Here $\gamma^2$ is the misfit
\begin{equation}
  \label{eq:misfit}
  \gamma^2 = \sum_i r_i^2
\end{equation}
which is expressed in terms of the residuals
\begin{equation}
  \label{eq:residual}
  r_i    = \frac{t_i - t^0_i}{\sigma}.
\end{equation}
They measure the deviation of the data points $t_i$ from the `true' values
$t_i^0$ (or the best estimates thereof) in units of the error bar $\sigma$.

For lack of further information, we argue that the likelihood 
\eqref{eq:gaussian_likelihood} is a reasonable choice and we note that it is,
in fact, the most uninformative probability distribution given only the mean
value and the variance and \emph{no further information}.\cite{sivia}
We want to stress that curve fitting with minimization of the misfit $\gamma^2$
is also implicitly based on the assumption of a Gaussian likelihood.

The problem of obtaining $\alpha^2F(\omega)$ from
$\tau^{-1}_{ex}(\omega)$ is extremely ill conditioned. This implies
that a direct solution will be totally dominated by noise and,
therefore, be completely meaningless. For this reason, all methods
discussed here involve a \emph{nuisance} or
regularization parameter that can be tuned in order to suppress noise
contributions. Apart from ad-hoc settings, a sensible choice is to adjust the
regularization parameter such that $\gamma^2 = N$ is obtained.
(See Appendix~\ref{app:A}.)

The first method of inversion is based on the relationship\cite{mars1}
\begin{equation}
  \label{eq:2}
  W(\omega) = \frac{1}{2\pi}\frac{d^2}{d\omega^2}\left[
  \omega\tau^{-1}_{op}(\omega)\right]
\end{equation}
which is approximately equal to $\alpha^2F(\omega)$
in the normal state at zero temperature. Application of this
formula to experimental data will result in numerical difficulties
because we have to keep in mind that the experimental signal
$\tau^{-1}_{ex}(\omega)$ consists, according to Eq.~\eqref{eq:3},
of at least three contributions. Two of these, namely $B(\omega)$
and $\eta(\omega)$ can obscure completely
the looked for spectral function $\alpha^2F(\omega)$ when the
second derivative of $\tau^{-1}_{ex}(\omega)$ is calculated.

On first sight, Eq.~\eqref{eq:2} would require that
$\tau^{-1}_{ex}(\omega)$ must be ambiguously smoothed `by hand'%
\cite{dorde} which is certainly not true. First of all, it is
much better to `smooth' the function $\omega\tau^{-1}_{ex}(\omega)$
which is monotonically increasing, much less structured, and
equal to zero at $\omega=0$. The application of standard
data processing techniques like Fast-Fourier-Transform (FFT)
smoothing or FFT low pass filters on this function allows one to
remove quite reliably the noise contribution $\eta(\omega)$.
For instance, the upper frequency threshold applied to the
FFT low pass filter will play the role of a nuisance (or
renormalization) parameter in this particular case. If
there is further knowledge about the background function $B(\omega)$
application of Eq.~\eqref{eq:2} is much safer than it looks on
first sight. We will discuss caveats and merits of this
{\it second derivative method} later on using computer generated
results which ensure $B(\omega)=0$ and which allow a controlled
 noise contribution $\eta(\omega)$.

The second method of inversion is based on the deconvolution
of the approximate relation
\begin{equation}
  \label{eq:4}
  \tau^{-1}_{op}(\omega;T) = \int\limits_0^\infty\!d\Omega\,
  K(\omega,\Omega;T)\alpha^2F(\omega),
\end{equation}
where $T$ denotes the temperature. The kernel $K(\omega,\Omega;T)$
is determined from theory. The caveat of this method is that
solutions of Eq.~\eqref{eq:4} for $\alpha^2F(\omega)$ are not
unique because, generally, the deconvolution of Eq.~\eqref{eq:4}
constitutes an ill-posed problem.

There are two approaches to solve this deconvolution problem and both
are based on a discretization of Eq.~\eqref{eq:4} of the form
\begin{equation}
  \label{eq:5}
 \tau^{-1}_{op}(\omega_i;T) = \sum\limits_{j=1}^{N_2}
 \Delta\Omega_j K(\omega_i,\Omega_j;T)\alpha^2F(\Omega_j), 
\end{equation}
with $i=1,\ldots,N_1$ and $\Delta\Omega_j = \Omega_{j+1}-\Omega_j$.
The first approach is straight forward and is called
{\it Singular Value Decomposition\/}\cite{nash} (SVD) which
is based on the vector form of Eq.~\eqref{eq:5}, namely
\begin{equation}
  \label{eq:6}
  {\bf t} = \mbox{\boldmath $K$}\,{\bf a},
\end{equation}
with the vector ${\bf t} = \{t_i=\tau^{-1}_{op,ex}(\omega_i;T)%
\vert i=1,\ldots,N_1\}$, the matrix $\mbox{\boldmath $K$} =
 \{K_{ij} = \Delta\Omega_jK(\omega_i,\Omega_j;T)\vert
i=1,\ldots N_1,\,j=1,\ldots N_2\}$, and the vector
${\bf a} = \{a_j = \alpha^2F(\Omega_j)\vert j=1,\ldots,N_2\}$.
Using SVD, the matrix \mbox{\boldmath $K$} of dimension
$N_1\times N_2$ is transformed into the matrix product
$\mbox{\boldmath $U$}\mbox{\boldmath $S$}\mbox{\boldmath $V$}^T$,
with \mbox{\boldmath $U$} and \mbox{\boldmath $V$} being unitary
matrices of dimension $N_1\times N_2$ and $N_2\times N_2$
respectively. The matrix $\mbox{\boldmath $S$} = \text{diag}
\left\{ s_j\vert j=1,\ldots, N_2\right\}$ with $s_j$ the
singular values (svs). Finally, $\mbox{\boldmath $V$}^T$ denotes
the transposed matrix \mbox{\boldmath $V$}. If the vector {\bf t}
and the matrix \mbox{\boldmath $K$} are known the vector {\bf a}
and, thus, $\alpha^2F(\omega)$ can be determined by `inverting'
Eq.~\eqref{eq:6}: ${\bf a} = \mbox{\boldmath $V$}\mbox{\boldmath $S$}'
\mbox{\boldmath $U$}^T\,{\bf t}$ with
$\mbox{\boldmath $S$}' = \text{diag}\{1/s_j\vert j=1,\ldots,N_2\}$.
However,  noise contained in the data {\bf t} will be dramatically
magnified by smallest svs, rendering the result meaningless. For this
reason, all contributions by svs below a certain threshold have to be
discarded by replacing the corresponding diagonal elements $1/s_i$
in the matrix $\mbox{\boldmath $S$}'$ by zeros.
This threshold plays the role of the nuisance parameter in the
SVD method. Dordevic {\it et al.}\cite{dorde} studied this approach
extensively and discussed in
particular the number of svs necessary to get a `smooth' spectral
function $\alpha^2F(\omega)$ together with a reasonable
reconstruction of the input data. In principle the problem of
`smoothing by hand' is moved from the input to the output of
the process. The caveat of this approach is the fact that it
doesn't ensure that the resulting spectral function $\alpha^2F(\omega)$
be positive definite. Most of the time, $\alpha^2F(\omega)$
will contain negative parts which cannot be removed even by applying
further regularization schemes.\cite{dorde}
Such negative parts are unphysical.

The second approach to the deconvolution problem is the so-called
{\em Maximum Entropy Method} (MaxEnt). 
Originally, E.T.~Jaynes\cite{jaynes} suggested the Maximum Entropy
principle for the assignment of probability distributions: If only
some testable information such as the mean value is given, one
should select that probability distribution $\{p_i\}$
which maximizes
the Shannon entropy\cite{sivia}
$S=-\sum_{i=1}^{N}p_i\log(p_i)$ subject to all
known constraints. In the case
where only the mean and the variance are known, the normal
distribution is the `most uninformative'
probability distribution (pdf). Although the `true' pdf may
be completely different, a normal distribution can be a sensible
choice for lack of further background information.

The MaxEnt principle has been generalized to the inference
of strictly positive functions such as the spectral function
$\alpha^2F(\omega)$ within Bayesian probability theory. This
fully probabilistic description allows for an explicit treatment
of the ambiguity inherent in badly conditioned problems and is
discussed in some detail in Appendix~\ref{app:A}.
In our particular case the generalized Shannon-Jaynes entropy
\eqref{eq:A3} is applicable with the default model vector {\bf m}
chosen to be constant. Most of the time this constant is chosen
in such a way that the spectral function $\alpha^2F(\omega)$
develops a certain high energy behavior.

The third method of inversion uses model spectral functions which depend on a
few parameters which are then determined using a {\it least
squares fit} to experiment based either on approximate formulas
of the form \eqref{eq:4} or full Eliashberg theory. Very often
preliminary results derived with the help of the second derivative
method from experiment (or using one of the other above mentioned
methods) can be utilized to minimize the number of parameters to
be fitted. Results from other experiments, for instance inelastic
neutron scattering etc., can easily be incorporated. Nevertheless,
in general this method will also result in non-unique solutions
for $\alpha^2F(\omega)$.

\subsection{Approximate Formulas}
\label{ssec:2.2}

For the normal state at zero temperature Allen\cite{allen}
provided a simplified form of the kernel of Eq.~\eqref{eq:4},
namely
\begin{equation}
  \label{eq:7}
  K(\omega,\Omega;T=0) = \frac{2\pi}{\omega}(\omega-\Omega)
  \theta(\omega-\Omega),
\end{equation}
where $\theta(x)$ is the step function. This formula is based on
a second order perturbation theory approach based on the weak
electron-phonon coupling in normal metals
and it is valid only in the clean limit, i.e.: no
impurities. To overcome the zero
temperature restriction Shulga {\it et al.}\cite{shulga1}
started from a full Eliashberg description of the electron-phonon
formalism and applied a series of approximations to reduce the
full results to the approximate form
\begin{eqnarray}
  \label{eq:8}
  K(\omega,\Omega;T) &=& \frac{\pi}{\omega}\left[
  2\omega\text{coth}\left(\frac{\Omega}{2T}\right)-(\omega+\Omega)
  \text{coth}\left(\frac{\omega+\Omega}{2T}\right)\right.\nonumber\\
  &&+\left.
  (\omega-\Omega)\text{coth}\left(\frac{\omega-\Omega}{2T}\right)
  \right],
\end{eqnarray}
which properly reduces to Eq.~\eqref{eq:7} for $T=0$. When applied
to invert data one has to keep in mind that this kernel becomes
singular for $\Omega=0$.

The work of Shulga {\it et al.} was generalized recently by
Sharapov and Carbotte\cite{sharap} to treat the possibility of a
pseudogap opening up in the fully dressed density of states
$\tilde{N}(\omega)$. They obtain
\begin{eqnarray}
  \label{eq:9}
  K(\omega,\Omega;T) &=& \frac{\pi}{\omega}
  \int\limits_{-\infty}^\infty\!d\varepsilon\,\left[
  \frac{\tilde{N}(\varepsilon-\Omega)}{N(0)}+
  \frac{\tilde{N}(\Omega-\varepsilon)}{N(0)}\right]\nonumber\\
  &&\times\left[n(\Omega;T)+f(\Omega-\varepsilon;T)\right]\left[
    f(\varepsilon-\omega;T)\right.\nonumber\\
  &&\left.-f(\varepsilon+\omega;T)\right],
\end{eqnarray}
which properly reduces to the result \eqref{eq:8} of Shulga
{\it et al.}\cite{shulga1} when $\tilde{N}(\omega)$ is taken
to be constant and equal to $N(0)$. Here $n(\omega;T)$ and
$f(\omega;T)$ are the Bose and Fermi distributions, respectively.
The zero temperature limit of Eq.~\eqref{eq:9} was obtained by
Mitrovi\'{c} and Fiorucci\cite{boza1} based on Allen's
second order perturbation theory approach.

Allen also provided a kernel similar to Eq.~\eqref{eq:7}
which applies approximately in the superconducting state at
zero temperature. In this case the kernel is of the form
\begin{eqnarray}
  \label{eq:10}
  K(\omega,\Omega;T=0) &=& \frac{2\pi}{\omega}(\omega-\Omega)
  \theta(\omega+2\Delta_0-\Omega)\nonumber\\
  &&\times
  E\left(\sqrt{1-\frac{4\Delta^2_0}{(\omega-\Omega)^2}}\right).
\end{eqnarray}
It ensures that $\tau^{-1}_{op}(\omega)$ is zero for
$0\le\omega\le 2\Delta_0$. Here, $E(x)$ is the complete elliptic
integral of the second kind and $\Delta_0$ is the energy gap
at $T=0$. It is valid in the clean limit only.
To derive Eq.~\eqref{eq:10} Allen treated the
superconducting transition within the framework of BCS theory,
i.e.: Eq.~\eqref{eq:10} is only valid for $s$-wave symmetry
of the superconducting order parameter. Moreover, $\Delta_0$
is an external parameter to Eq.~\eqref{eq:10} and its value
has to be determined by other means. Treating the superconducting
transition within the framework of Eliashberg theory will
certainly go beyond the possibilities of Eq.~\eqref{eq:10} and
this will have to be kept in mind when Eq.~\eqref{eq:10} is applied
to invert superconducting state optical data of real $s$-wave
superconductors which are well known to be exceptionally well
described by Eliashberg theory.\cite{carb1}

A consensus exists that in the high $T_c$ cuprates the superconducting
order parameter is of $d$-wave rather than $s$-wave symmetry.
Here we follow Carbotte and Schachinger\cite{carb2} and simulate
(in a first approximation) the effect of $d$-wave in the
Allen formula \eqref{eq:10} for $s$-wave by simply averaging it
over a distribution of gaps having $d$-wave symmetry.
The result is
that Eq.~\eqref{eq:10} needs to be averaged over the polar angle
$\vartheta$ of the two dimensional CuO$_2$ Brillouin zone. This
results in the kernel
\begin{eqnarray}
  \label{eq:11}
  K(\omega,\Omega;T=0) &=& \frac{2\pi}{\omega}\left\langle(\omega-\Omega)
  \theta(\omega+2\Delta_0(\vartheta)-\Omega)\right.\nonumber\\
  &&\left.\times
  E\left(\sqrt{1-\frac{4\Delta^2_0(\vartheta)}{(\omega-\Omega)^2}}\right)
  \right\rangle_\vartheta,
\end{eqnarray}
with $\langle\cdots\rangle_\vartheta$ denoting the
$\vartheta$-average which can be limited to the interval
$\vartheta\in[0,\pi/4]$ for symmetry reasons. Furthermore,
$\Delta_0(\vartheta) = \Delta_0\cos(2\vartheta)$ reflecting the
$d$-wave symmetry of the superconducting order parameter.
Eq.~\eqref{eq:11} ensures that the optical scattering rate is
finite in the superconducting state for $\omega>0$. This is in
contrast to what is observed in $s$-wave superconductors.

\section{Numerical Results}
\label{sec:3}
\subsection{Normal Metals}
\label{ssec:3.1}

\begin{figure}[tp]
  \vspace*{-7mm}
  \includegraphics[width=9cm]{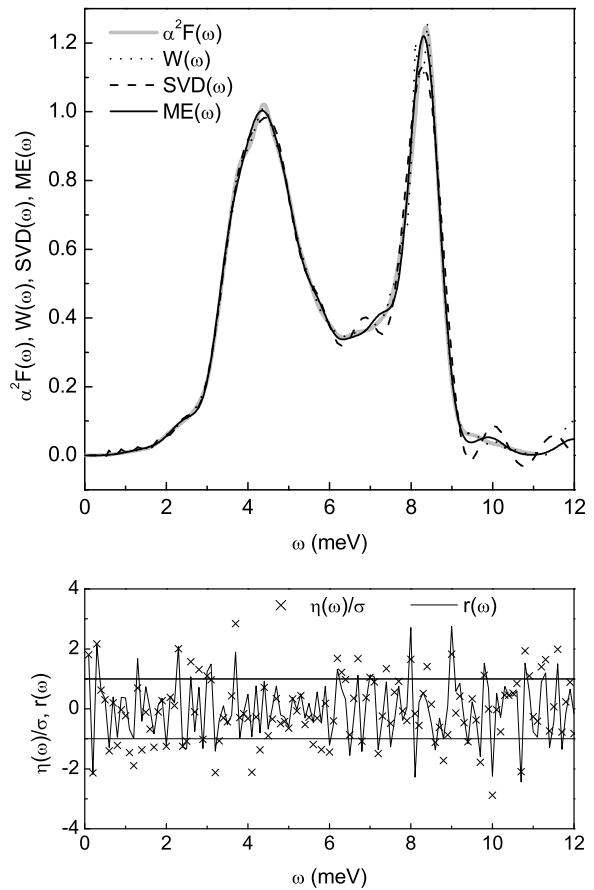}
  \caption{Top frame: Inversion of zero temperature, normal state
  optical scattering rate data $\tau^{-1}_{op}(\omega)$
  of lead computer generated using the kernel
  \protect{\eqref{eq:7}}. The gray solid line indicates the
  $\alpha^2F(\omega)$ employed to generate the data.
  The dotted line
  corresponds to the function $W(\omega)$ according to
  Eq.~\protect{\eqref{eq:2}}, SVD($\omega$) (dashed line) shows the
  result of an SVD inversion, and ME($\omega$) (solid line)
  presents the result of a MaxEnt inversion. Bottom frame: The
  crosses correspond to the normalized uncorrelated Gaussian noise
  $\eta(\omega)/\sigma$ which was added to the input
  data for the MaxEnt inversion and $r(\omega)$ (solid line) gives the
  residual of the MaxEnt data reconstruction.
}
  \label{fig:1}
\end{figure}
We will study in quite some detail various inversion techniques
using, as a first material, lead. The electron-phonon spectral
density $\alpha^2F(\omega)$ was derived from tunneling data by
McMillan and Rowell.\cite{mcmillan} This spectrum, which is represented
in the top frame of
Fig.~\ref{fig:1} by a gray solid line, has two distinctive peaks
which are separated from each other by about $4\,$meV. The Debye
energy $\omega_D = 11.2\,$meV. Optical data for lead was obtained
by Joyce and Richards\cite{joyce} and later by Farnworth and
Timusk.\cite{farnworth} The extracted $\alpha^2F_{tr}(\omega)$
was found to be in remarkable good agreement with earlier tunneling
data and with the results of direct band structure calculations
of $\alpha^2F(\omega)$ by Tomlinson and Carbotte.\cite{tomlinson}
In this subsection only computer
generated optical scattering rate data based on the various
kernels discussed in Sec.~\ref{ssec:2.2} and on complete Eliashberg
equations (Appendix~\ref{app:B}) will be used to study the
various inversion techniques. This provides us with data free
of a background signal $B(\omega)$ and with controlled noise
$\eta(\omega)$.

In a first step zero temperature, normal state
$\tau^{-1}_{op}(\omega)$ data are computer generated
using kernel \eqref{eq:7}. We calculate the function $W(\omega)$
(dotted line, upper frame Fig.~\ref{fig:1})
using the second derivative method
and the agreement with the input spectrum (gray solid line) is
almost perfect without the need of `smoothing by hand'.
(Only the second peak shows oscillations.) Inversion of the input
data using the SVD method results in the curve SVD($\omega$)
(dashed line, upper frame Fig.~\ref{fig:1}).
The svs threshold was set to $10^{-3}$, i.e.: 87 svs have been used. A few
wiggles remain in the valley between the two peaks and we see oscillations
at energies $> 10\,$meV which also go negative. Data beyond the
Debye energy are irrelevant.

For the application of the MaxEnt method we add uncorrelated
Gaussian noise of standard deviation $\sigma=10^{-3}$ to ensure
a controlled error distribution for the computer generated
$\tau_{op}^{-1}(\omega)$ data.
(This, in principle, biases the comparison in favor of the
second derivative and SVD method.) The curve ME($\omega$)
(solid line, upper frame Fig.~\ref{fig:1}) presents the result of the
MaxEnt inversion in which we used optimized preblur
[blur-width $b=0.05$, see Eq.~\eqref{eq:preblur}]
and the default model was set to $m_j=0.001$.
ME($\omega$) underestimates slightly the second
peak but otherwise shows perfect agreement with the input spectrum.
We also see an additional feature beyond the Debye energy which is
irrelevant because it reflects the default model.
The bottom frame of Fig.~\ref{fig:1} demonstrates the
quality of the data reconstruction achieved by the MaxEnt method.
The crosses symbolize the normalized noise $\eta(\omega)/\sigma$
which was added to the optical
scattering rate and the solid line corresponds to the residual
\eqref{eq:residual}. As we
only added noise to the computer generated $\tau^{-1}_{op}(\omega)$,
$r(\omega)$ should track the normalized noise $\eta(\omega)/\sigma$,
as it does.

\begin{figure}[tp]
  \vspace*{-2mm}
  \includegraphics[width=9.5cm]{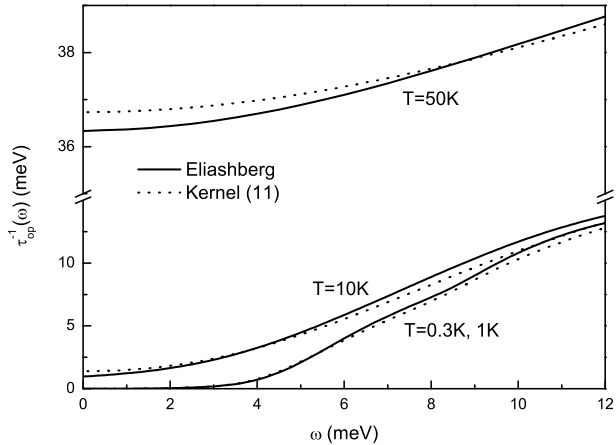}
  \caption{Temperature dependence of the normal state optical scattering
rate $\tau^{-1}_{op}(\omega)$ of lead. The solid lines correspond to results
derived using Eliashberg theory and the dotted lines correspond to
data generated using kernel \protect{\eqref{eq:8}}. Temperatures are
$0.3\,$K, $1\,$K, $10\,$K, and $50\,$K.
}
  \label{fig:2}
\end{figure}
Zero temperature is not a realistic case and we proceed to study
normal state, finite temperature results. There are two options to
computer generate $\tau^{-1}_{op}(\omega)$ data:
(a) kernel \eqref{eq:8} is
applied, or (b) Eliashberg theory (see Appendix~\ref{app:B}) is
used.
The superconducting order parameter is zero in the normal state and
the renormalization formula \eqref{eq:B2} takes on a closed form.
Fig.~\ref{fig:2} presents our results for the temperature dependence
of the optical scattering rate in lead for four different temperatures,
namely $0.3\,$K, $1\,$K, $10\,$K, and $50\,$K. The results according
to Eliashberg theory are presented by solid lines, while the dotted
lines correspond to the results of kernel \eqref{eq:8}. There are
small but distinct differences between the two sets of data.

\begin{figure}[tp]
  \includegraphics[width=9cm]{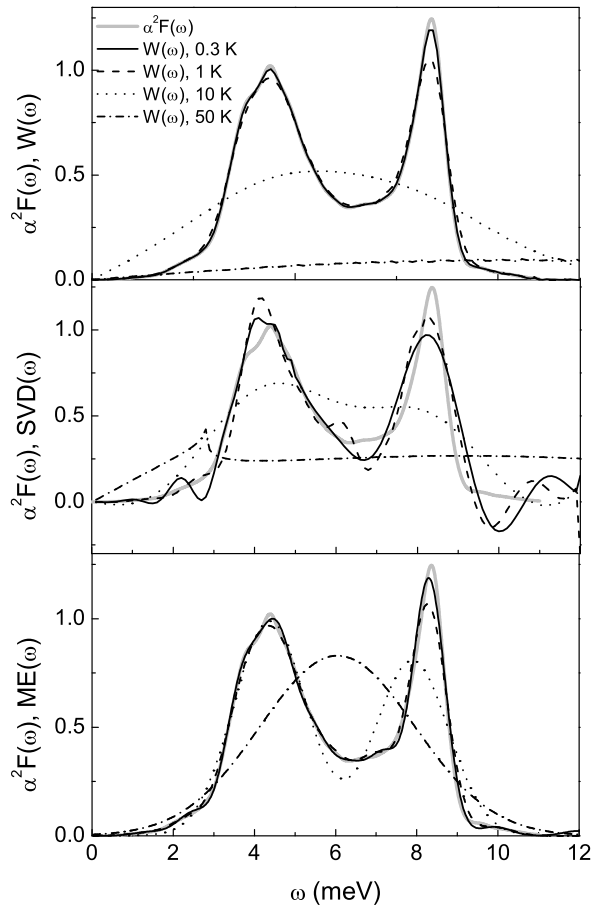}
  \caption{Inversion of finite temperature, normal state
$\tau^{-1}_{op}(\omega)$ data computer generated using kernel
\protect{\eqref{eq:8}}.
The solid lines correspond to the temperature $T=0.3\,$K, dashed
lines to $1\,$K, dotted lines to $10\,$K, and dash-dotted lines
to $50\,$K. The gray solid line represents the $\alpha^2F(\omega)$
spectral function applied to calculate the optical scattering rate
data. Top frame: Second derivative method. Center frame:
SVD method. Bottom frame: MaxEnt method.}
  \label{fig:3}
\end{figure}
Fig.~\ref{fig:3} presents the spectra $W(\omega)$, SVD($\omega$), and
ME($\omega$) which result from the application of the first two
methods of inversion discussed in Sec.~\ref{ssec:2.1}. As input
we used the finite temperature normal state optical scattering rate
for lead generated using kernel \eqref{eq:8} (dashed lines of
Fig.~\ref{fig:2}).
The top frame presents as a result of the second derivative method,
the function $W(\omega)$ as defined in Eq.~\eqref{eq:2}. At the
lowest temperature, $T=0.3\,$K the input $\alpha^2F(\omega)$
(grey solid line) is
perfectly reproduced (solid line), while at $T=1\,$K (dashed line)
the high energy peak is already underestimated. At $T=10\,$K
(dotted line) the method is no longer able to resolve the two
peak structure, and at $T=50\,$K (dash-dotted line) the method
fails completely. Nevertheless, it has to be emphasized that no
smoothing had to be applied to the input data as no artificial noise
was added.

The center frame of Fig.~\ref{fig:3} presents the results SVD($\omega$)
of a singular value decomposition of Eq.~\eqref{eq:6} using kernel
\eqref{eq:8}. The svs threshold was
set to $10^{-3}$. At $T=0.3\,$K and $1\,$K we obtain
reasonable agreement
with the $\alpha^2F(\omega)$. For energies $> 9\,$meV the inversion
shows oscillations and SVD($\omega$) even becomes negative which is
unphysical. We also see oscillations at low energies and between the
two peaks. At $T=10\,$K the SVD method still resolves a hint of
a two peak structure in contrast to the second derivative method.
Finally, at $T=50\,$K the method fails.

The bottom frame of Fig.~\ref{fig:3} presents the results ME($\omega$)
of the MaxEnt deconvolution of Eq.~\eqref{eq:5}.
Uncorrelated Gaussian noise of $\sigma=10^{-3}$ was added
to the computer generated data. For the inversion optimized
preblur (see Appendix~\ref{app:A}) was applied with $b=0.4$ for $T=0.3\,$K,
$b=0.46$ for $1\,$K, $b=0.89$ for $T=10\,$K,
and $b=1.91$ for $T=50\,$K. The default model was set to $m_j=0.01$.
The $T=0.3\,$K inversion (solid line) gives almost perfect agreement
with the model $\alpha^2F(\omega)$ spectral function. At
$T=1\,$K the high energy peak is underestimated but reproduced at the
appropriate energy. At $T=10\,$K,
the two peak structure is still well resolved, only the second peak is
underestimated and shifted to lower energies. The result is certainly
much better than that of the other two methods. Finally, at $T=50\,$K
MaxEnt is no longer able to resolve the two peak structure. Nevertheless,
it is quite interesting to note that the area under the dash-dotted
curve is $3.95\,$meV which is very close to the area $4.03\,$meV under
the original $\alpha^2F(\omega)$ spectral function.

\begin{figure}[tp]
  \vspace*{-7mm}
  \includegraphics[width=9cm]{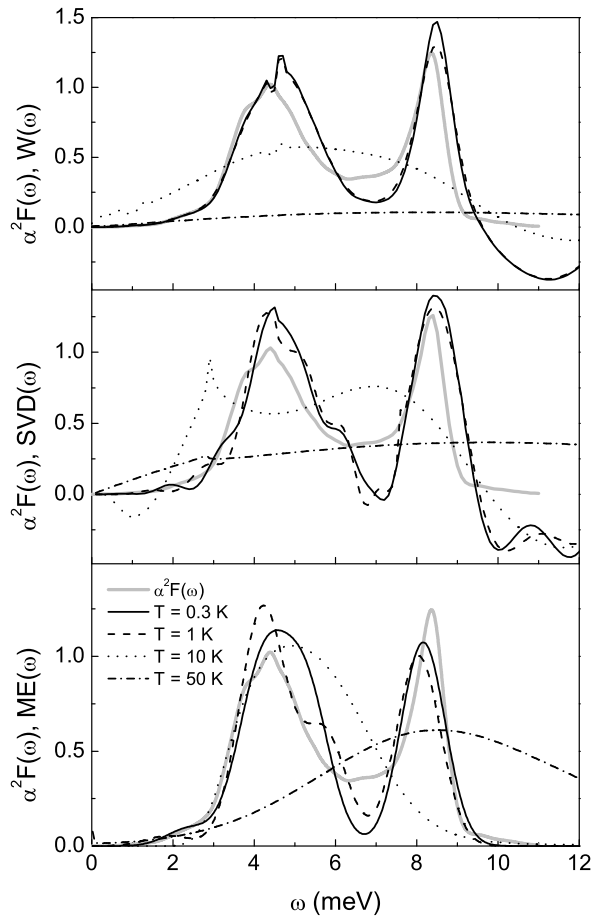}
  \caption{The same as Fig.~\protect{\ref{fig:3}} but now the Eliashberg
theory generated normal state, finite temperature $\tau^{-1}_{op}(\omega)$
data (solid lines in Fig.~\protect{\ref{fig:2}}) are used as input.}
  \label{fig:4}
\end{figure}
Fig.~\ref{fig:2} demonstrated that $\tau^{-1}_{op}(\omega)$ data
computer generated
from full normal state Eliashberg theory differ from the approximate
results of kernel \eqref{eq:8}. It can also be assumed that
metals will more likely follow the predictions of full Eliashberg
theory rather than approximate model formulas. It is therefore
interesting to investigate how the inversion on the basis of the
approximate kernel \eqref{eq:8} performs when $\tau^{-1}_{op}(\omega)$
data computer generated within full Eliashberg theory are used
for inversion. The
result is presented in Fig.~\ref{fig:4} which is organized the same
way as Fig.~\ref{fig:3}. The top frame demonstrates the application
of the second derivative formula which shouldn't have any problems
because this method is not based on approximate models formulas.
Nevertheless, $W(\omega)$ is only in reasonable
agreement with the original $\alpha^2F(\omega)$ at low temperatures.
The low energy peak is over estimated and shifted towards higher
energies, the valley between the peaks is too low, the second peak
is positioned at the correct energy but its height is over/underestimated.
At $T=10\,$K the two peak structure is no longer resolved and
at $T=50\,$K the inversion fails. The central frame of Fig.~\ref{fig:4}
presents the function SVD($\omega$) as a result of an SVD inversion.
The svs threshold was set at $10^{-2}$. At low temperatures, the
peak positions are at the proper energies, nevertheless the low
energy peak is overestimated, the valley between the peaks underestimated,
and the high energy peak is too wide.
At $T=10\,$K a two peak structure is resolved but both peaks are
placed at the wrong energies. At $T=50\,$K the method fails altogether.
It is typical for this method to show oscillations at energies
$> 9\,$meV which result in unphysical negative contributions even at
energies below $\omega_D$.

Before applying the MaxEnt inversion uncorrelated Gaussian noise of
$\sigma=0.1$ was added to the input data. For the temperature $0.3\,$K and
$1\,$K the preblur parameter was optimized to 0.54 at higher temperatures
no preblur was applied.
The default model was set to $0.01$. At low temperatures MaxEnt
overestimates the low energy peak and/or makes it broader.
The valley between the peaks is too low, the second peak is
resolved reasonably well and is at the proper position but
underestimated in height. At higher
temperatures the two peak structure is no longer resolved ($T=10\,$K,
dotted line and $T=50\,$K, dash-dotted line). Nevertheless, data
reconstruction is within error bars and this proves that we face in
this case a deconvolution problem which is particularly ill
conditioned.

All this demonstrates quite clearly that the application of methods of
inversion based on approximate models to experimental data
(represented here by computer generated $\tau^{-1}_{op}(\omega)$
data using full Eliashberg theory) can
quite easily result in deconvoluted spectra $\alpha^2F(\omega)$ which
will be close but not necessarily equal to the real electron-phonon
spectrum which governs the interaction. In particular, the deviations
from the gray solid lines in the central and bottom frame of
Fig.~\ref{fig:4} represent the deviations from the `real'
$\alpha^2F(\omega)$ which are required by the approximate
kernels to reproduce the input optical scattering rate data as well as
possible.

\begin{figure}[tp]
  \vspace*{-7mm}
  \includegraphics[width=9cm]{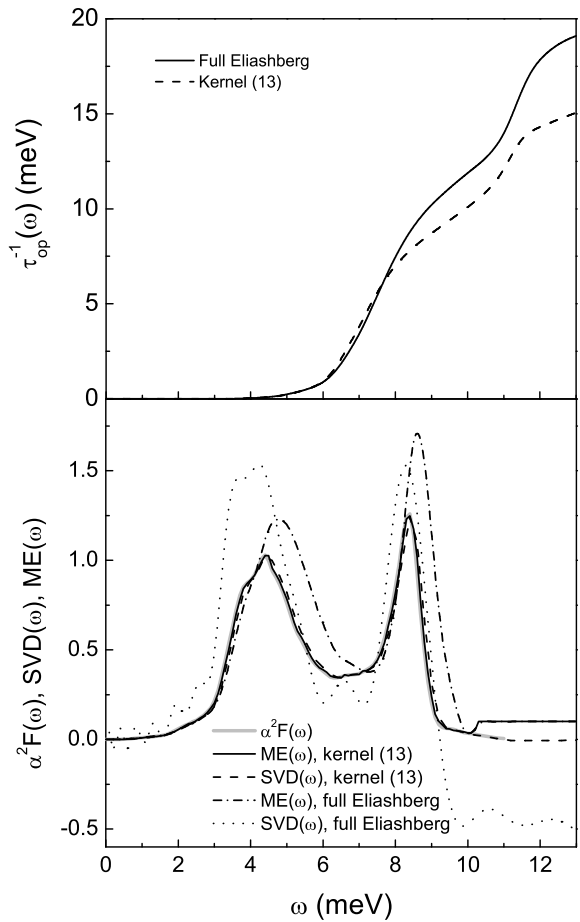}
  \caption{Top frame: The computer generated
superconducting state optical scattering rate
$\tau^{-1}_{op}(\omega)$ for Pb at
$T=0.05T_c$ with $T_c = 7.2 K$. The solid line is based on the
full Eqs.~\protect{\eqref{eq:Bx}} while the dashed line is obtained
using the simplified kernel \protect{\eqref{eq:10}}. Bottom frame:
The gray solid line symbolizes the $\alpha^2F(\omega)$ of Pb. The solid
line is the result of the inversion of $\tau_{op}^{-1}(\omega)$ data
computer generated using kernel \protect{\eqref{eq:10}} (dashed line
in the top frame of this figure) using the
MaxEnt method while the dashed line corresponds to
an SVD inversion. The dash-dotted and dotted lines present equivalent
results but now for full Eliashberg data (solid line in the top
frame of this figure).}
  \label{fig:5}
\end{figure}
We now move on to a discussion of superconducting state data. The
top frame of Fig.~\ref{fig:5} presents the results for the
superconducting state optical scattering rate $\tau^{-1}_{op}(\omega)$
in Pb at $T=0.05T_c$ with $T_c = 7.2\,$K. The solid line was obtained
on evaluation of the full Eqs.~\eqref{eq:Bx} and \eqref{eq:B4} taken
for $s$-wave symmetry of the superconducting order parameter and
a Coulomb pseudopotential $\mu^\ast=0.1438$. The dashed line is for
comparison and was obtained from kernel \eqref{eq:10} using the
electron-phonon spectral density $\alpha^2F(\omega)$ shown
by a gray solid line in the bottom frame of this figure. The
approximate kernel \eqref{eq:10} is evaluated with
$\Delta_0 = 1.39\,$meV, the gap edge predicted by the full
Eliashberg calculation.

The bottom frame of Fig.~\ref{fig:5} presents the result of SVD
as well as MaxEnt inversions based on the approximate kernel
\eqref{eq:10}. The dashed line corresponds to the
SVD inversion (svs threshold was set at $10^{-2}$) of the optical
scattering rate generated using kernel \eqref{eq:10} (dashed line
in the top frame of Fig.~\ref{fig:5}) and $\Delta_0 = 1.39\,$meV.
As expected, the agreement
is almost perfect. The dotted line, on the other hand, shows the
result of an SVD inversion of full Eliashberg data (solid line
in the top frame). Here, the lower, transverse peak centered around
$\sim 4\,$meV is broader and the area under the peak is larger.
The same holds for the upper, longitudinal phonon peak but the
differences are now less pronounced. Beyond $\sim 9\,$meV the
dotted line becomes negative giving unphysical results. These
differences are due to the use of the approximate kernel
\eqref{eq:10} in the inversion process.

The MaxEnt inversion was performed by attaching error bars of
$\sigma = 10^{-2}$ to the data and by adding uncorrelated
Gaussian noise of the same $\sigma$.
Furthermore, no preblur was applied and the default model was set to
0.1. The solid line is the MaxEnt inversion of the
data computer generated with the help of kernel \eqref{eq:10}
(dashed line in the top frame of Fig.~\ref{fig:5}). Again we achieve
perfect agreement. At energies $> 10.5\,$meV the function ME($\omega$)
levels off at the value 0.1 demonstrating the influence of the
chosen default model. The dash-dotted curve, on the other hand, is
based on full Eliashberg theory generated input data (solid line
in the top frame of this figure). Both peaks are now overestimated
in their height and width and are shifted towards higher
energies. Nevertheless, ME($\omega$) never becomes negative which
proves that there exists a positive definite solution for the
deconvolution problem of Eq.~\eqref{eq:4}. As experimental
data are more
likely to be close to full Eliashberg theory results, the
deconvolution of Eq.~\eqref{eq:4} on  the basis of kernel \eqref{eq:10}
will result in an electron-phonon spectral density $\alpha^2F(\omega)$
which will not agree in all details with the real spectral density
despite the fact that the input data will be excellently reproduced.
The only possible check for the validity of the deconvoluted
electron-phonon spectral density SVD($\omega$) or ME($\omega$) is
using it to calculate the optical scattering rate based on
full Eliashberg theory and compare with the data. Such a comparison
will then result in necessary readjustments of the deconvoluted
spectrum.

\subsection{High-$Tc$ cuprates}
\label{ssec:3.2}

In contrast to the normal metal lead, the high $T_c$ cuprates are
not likely to be electron-phonon systems, they are known to be
highly correlated systems. There is a class of models used to
describe such systems which we will refer to as boson
exchange models. They have many common elements with the
electron-phonon case. In particular, there exists a well
developed literature on the Nearly Antiferromagnetic Fermi Liquid
model (NAFFL) introduced by Pines and collaborators.\cite{pines1,%
pines2} The exchange bosons are antiferromagnetic spin fluctuations
as described by Millis {\it et al.}\cite{mmp} (MMP).
Within this model the Eliashberg equations are retained
in a zeroth order approximation neglecting possible vertex
corrections which go beyond Migdal's theorem.
The electron-phonon spectral density $\alpha^2F(\omega)$
is replaced by the imaginary part of the spin susceptibility
multiplied by the square of a coupling of the spin fluctuations
to the charge carriers. In general, this interaction is anisotropic
and not pinned to the Fermi surface.\cite{branch} Nevertheless,
as a first approximation, one can work with a simple interaction
spectral function $I^2\chi(\omega)$ which replaces the
$\alpha^2F(\omega)$ of Eliashberg theory. Carbotte {\it et al.}%
\cite{carb3} found that in optimally doped, twinned
YBCO single crystals the measured normal state
optical scattering rate $\tau^{-1}_{ex}(\omega)$, reported by
Basov {\it et al.}\cite{bas1}, can be well
described by a single MMP form:
\begin{equation}
  \label{eq:12}
  I^2\chi(\omega) = I^2\frac{\omega/\omega_{SF}}
  {1+(\omega/\omega_{SF})^2}.
\end{equation}
The two parameters, the square of the coupling constant $I$ and the
characteristic spin fluctuation energy $\omega_{SF}$ were
determined from a least squares fit to the data in the energy
interval $0\le\omega\le 250\,$meV. The values
$I^2 = 0.83$ and $\omega_{SF}=20\,$meV have been reported by CSB.

\begin{figure}[tp]
  \vspace{-7mm}
  \includegraphics[width=9cm]{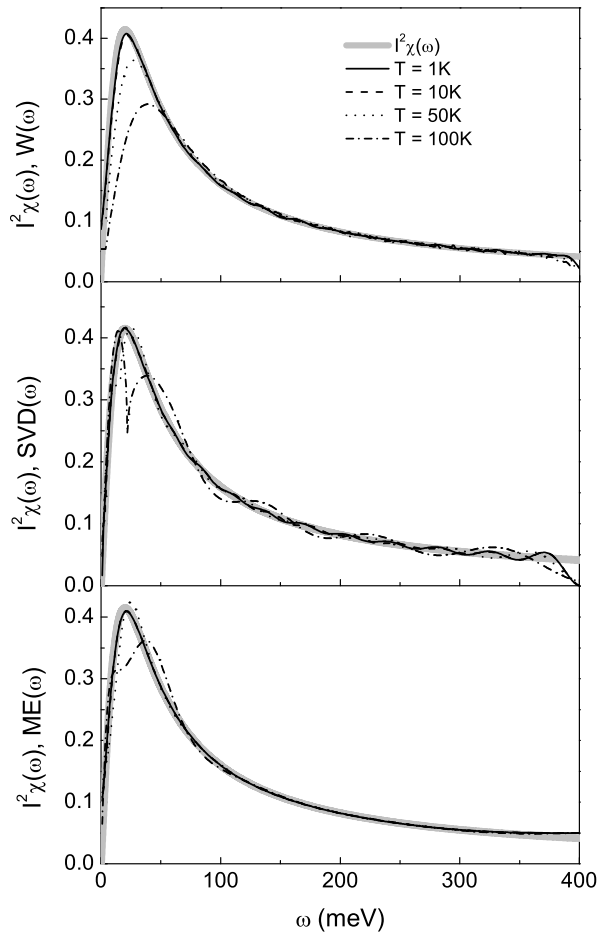}
  \caption{The same as Fig.~\protect{\ref{fig:3}} with the
$\alpha^2F(\omega)$ replaced by the $I^2\chi(\omega)$ defined by
Eq.~\protect{\eqref{eq:12}} with $I^2=0.83$ and $\omega_{SF}=20\,$meV.
The solid lines corresponds to $T=1\,$K, the dashed lines to
$10\,$K, the dotted lines to $50\,$K, and the dash-dotted lines
to $100\,$K.}
  \label{fig:6}
\end{figure}

\begin{figure}[tp]
  \vspace*{-7mm}
  \includegraphics[width=9cm]{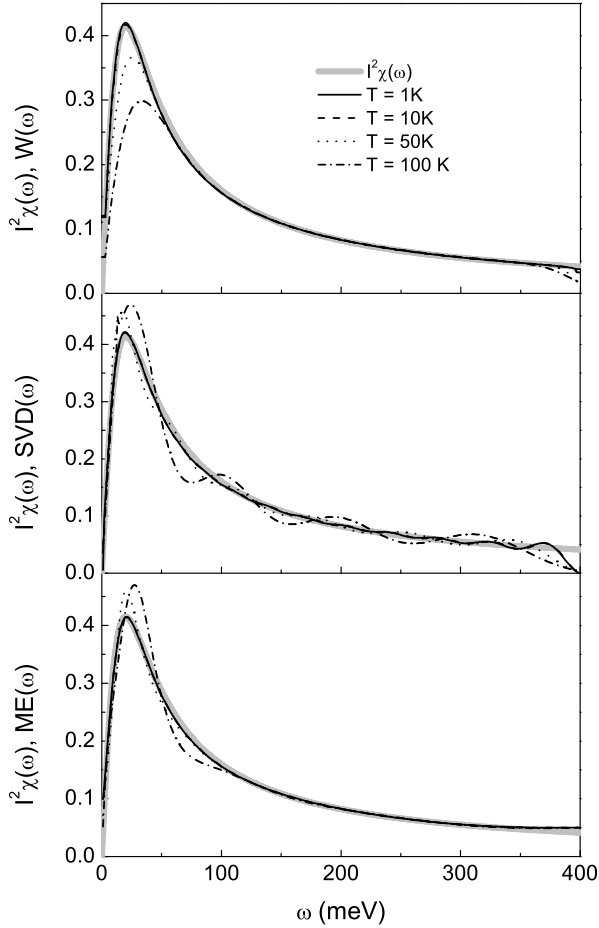}
  \caption{The same as Fig.~\protect{\ref{fig:6}} but now
Eliashberg theory generated optical scattering rates have
been used as input for the inversion.}
  \label{fig:7}
\end{figure}

Based on the results for lead we cannot necessarily expect
inversions of $\tau^{-1}_{op}(\omega)$ measured around $T=100\,$K or
even higher will be feasible. To investigate this, normal state
$\tau^{-1}_{op}(\omega)$ data are computer generated
at various temperatures, namely $T=1\,$K, $10\,$K,
$50\,$K, and $100\,$K using either kernel \eqref{eq:8} or full
Eliashberg theory with an $I^2\chi(\omega)$ determined by
Eq.~\eqref{eq:12} and with the above values for the
parameters $I^2$ and $\omega_{SF}$.
The results of the inversion based on
the approximate kernel \eqref{eq:8} are discussed in Fig.~\ref{fig:6}
with $\tau^{-1}_{op}(\omega)$ data generated using
kernel \eqref{eq:8} and, in Fig.~\ref{fig:7} with
$\tau^{-1}_{op}(\omega)$ generated by full Eliashberg theory.

The top frame of Fig.~\ref{fig:6} presents results for $W(\omega)$
from the second derivative method. At the two lowest
temperatures, namely $T=1\,$K (solid line) and $10\,$K (dashed line)
the $I^2\chi(\omega)$
spectrum (gray solid line) is almost perfectly reproduced.
At higher temperatures, namely at $T=50\,$K (dotted line) and
$100\,$K (dash-dotted line) the inverted spectrum develops a
less pronounced peak which is also shifted towards higher energies.
In the tail ($\omega>100\,$meV) noise develops in the inverted
spectrum. Nevertheless, in contrast to lead with its narrow two
peak structure the simple MMP form can easily be inverted from
optical scattering rate data even at temperatures around
$100\,$K.

The center frame of Fig.~\ref{fig:6} presents the results
SVD($\omega$) of a singular value decomposition. The svs
threshold was set at $10^{-3}$ for $T=1\,$K and $10\,$K and was increased
to $10^{-2}$ for $T=50\,$K and $100\,$K. The inverted spectrum
agrees reasonably well with the original spectrum at low
energies, $\omega <75\,$meV. At higher energies significant
oscillations occur. We also note for $T=100\,$K a typical splitting
of the peak at $20\,$meV into two peaks. This is an indication
of a particularly ill conditioned inversion problem. This
phenomenon can be observed for rather narrow and fast rising peaks.%
\cite{sivia}

The bottom frame of Fig.~\ref{fig:6} presents the results
ME($\omega$) of a MaxEnt deconvolution. The error bars on the
data were determined by $\sigma=0.15$ and uncorrelated Gaussian
noise of the same $\sigma$ was added. No preblur was applied.
The default model was set to 0.05. This ensures that the spectrum
ME($\omega$) agrees at high energies with $I^2\chi(\omega)$.
The agreement with the original $I^2\chi(\omega)$ spectral
function (gray solid line) is excellent up to temperatures of
$50\,$K. At $100\,$K the peak
at $20\,$meV is not as well resolved and shows tendency towards a
splitted peak similar to the SVD inversion at the same temperature.
Otherwise, the agreement is still rather good.

\begin{figure}[tp]
  \vspace*{-7mm}
  \includegraphics[width=9cm]{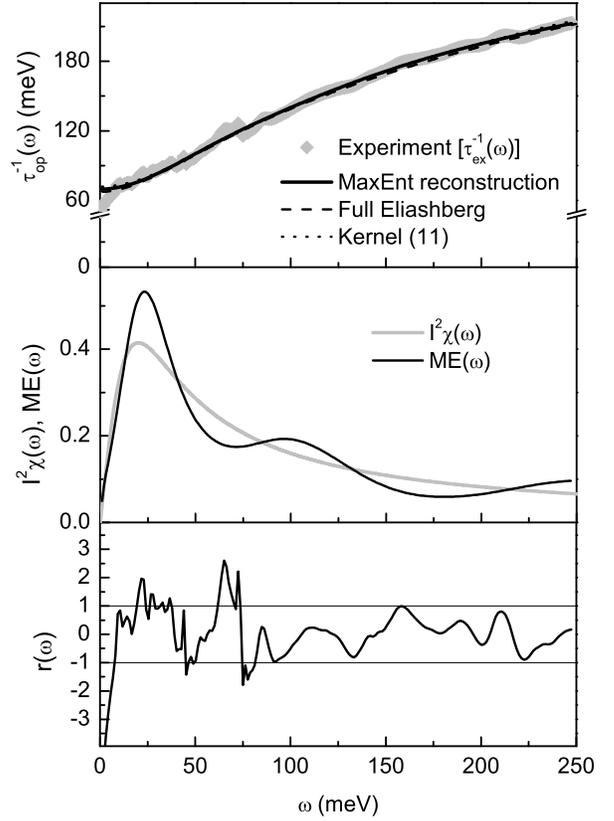}
  \caption{Top frame: 
Experimental normal state optical scattering rate
$\tau^{-1}_{ex}(\omega)$ at $T=95\,$K for an optimally doped,
twinned YBCO$_{6.95}$ single crystal as reported
by Basov {\it et al.}\protect{\cite{bas1}} (solid gray diamonds).
The black solid line corresponds to the MaxEnt reconstruction based
on kernel \protect{\eqref{eq:8}} using the spectral function
ME($\omega$) shown in the middle frame of this figure (black solid line).
The dashed line presents $\tau^{-1}_{op}(\omega)$
of a full Eliashberg calculation
based on the spectral function $I^2\chi(\omega)$ shown by a gray solid
line in the middle frame of this figure and the dotted line
corresponds to $\tau^{-1}_{op}(\omega)$ computer generated using kernel
\protect{\eqref{eq:8}} and the same $I^2\chi(\omega)$.
Bottom frame: The residual
$r(\omega)$ according to Eq.~\protect{\eqref{eq:11}} of the MaxEnt
data reconstruction.
}
  \label{fig:8}
\end{figure}

The results presented in Fig.~\ref{fig:7} are very similar to the ones
shown in Fig.~\ref{fig:6}
with the difference that SVD and MaxEnt (error bars were determined
by $\sigma=0.2$, uncorrelated Gaussian noise
of the same $\sigma$ was added, no preblur)
now overestimate the
peak at $20\,$meV. The SVD result for $100\,$K (dash-dotted
line in the center frame of Fig.~\ref{fig:7}) develops a slight
tendency for a splitted peak at $20\,$meV. All this
is the result of the minor differences between
the optical scattering rates calculated from kernel \eqref{eq:8}
and full Eliashberg theory and the top frame of Fig.~\ref{fig:8}
demonstrates how little these differences are.

We proceed to a study of the MaxEnt inversion of experimental data
and make use of the $T=95\,$K normal state optical scattering
rate measured by Basov {\it et al.}\cite{bas1} on an optimally doped,
twinned YBCO$_{6.95}$ single crystal. Fig.~\ref{fig:8} presents the result
of a MaxEnt inversion. We assume the experimental data
$\tau^{-1}_{ex}(\omega)$ to be
contaminated by a substantial uncorrelated Gaussian noise of $\sigma=2.5$.
The inversion is performed on the basis of kernel \eqref{eq:8} and
the resulting spectral function ME($\omega$) is shown as a solid
line in the middle frame of Fig.~\ref{fig:8}. ME($\omega$) then
replaces $\alpha^2F(\omega)$ in Eq.~\eqref{eq:4} which is used
to calculate the
reconstructed optical scattering rate shown by a solid line
in the top frame of Fig.~\ref{fig:8}. It reproduces excellently
the experimental data (gray solid diamonds). For comparison,
the top frame of this figure contains
two more results, namely $\tau^{-1}_{op}(\omega)$ calculated
from full Eliashberg theory (dashed line) using the $I^2\chi(\omega)$
reported by CSB, namely $I^2 = 0.83$ and $\omega_{SF} = 20\,$meV
(gray solid line in the center frame of Fig.~\ref{fig:8}).
The dotted line, on the other hand, corresponds to
$\tau^{-1}_{op}(\omega)$ calculated from Eq.~\eqref{eq:4} using
kernel \eqref{eq:8} and the same $I^2\chi(\omega)$. Obviously,
the two results are very close with the dotted line slightly
above the dashed one at low energies. The opposite holds for high
energies. This is in agreement with the result found for lead
(see Fig.~\ref{fig:2}).

Finally, the bottom frame of Fig.~\ref{fig:8} shows the residual
$r(\omega)$ which is a measure for the quality of the data
reconstruction. Apart from the low energy region the reconstruction
is within the assumed standard deviation (indicated by the
two straight lines at 1 and -1) of $\sigma=2.5$. The result is
to be compared with the $r(\omega)$ shown in the bottom frame
of Fig.~\ref{fig:1} which results from the reconstruction of
computer generated data with additional uncorrelated Gaussian noise.
The $r(\omega)$ in the bottom frame of Fig.~\ref{fig:8} is a rather
smooth function which contains for energies $> 70\,$meV very little
stochastic elements which could be identified as noise. The various
data points appear to be rather correlated an effect which could
either be attributed to an additional background function
$B(\omega)$ [see Eq.~\eqref{eq:3}] or to a `real' signal. Nevertheless,
what is important here is the fact that the inverted
ME($\omega$) has a nonzero contribution even at energies
$> 150\,$meV thus establishing a high energy background in
$I^2\chi(\omega)$ as predicted by CSB. Finally, as both spectral
functions presented in the central frame of Fig.~\ref{fig:8}
reconstruct the experimental
data (solid gray diamonds in the top frame of Fig.~\ref{fig:8})
equally well, they
can be used as valid spectral functions because of the non-uniqueness
of the deconvolution problem. Further calculations and comparison
with other experiments than optical conductivity may then help
to discriminate between these two spectra. It is interesting
to point out that the area under $I^2\chi(\omega)$ ($42\,$meV)
is approximately reproduced by the area under the spectrum ME($\omega$)
($41.4\,$meV) which could be used to explain the
oscillations in ME($\omega$) as a result of the enhanced main peak
at $20\,$meV.

Tu {\it et al.}\cite{tu} measured the optical scattering rate
of optimally doped Bi$_2$Sr$_2$CaCu$_2$O$_{8+\delta}$ (Bi2212)
single crystals at various temperatures. They derived, using data
analysis different from the methods discussed here, that even
in the normal state at $100\,$K a resonance peak is seen in the
function $W(\omega)$ while it is rather featureless at $295\,$K.
Schachinger and Carbotte\cite{schach1} also analyzed these data
using a combination of the second derivative method and least squares
fits to the data. In particular, they found that the $T=295\,$K
data are well described by an MMP form \eqref{eq:12} in the
energy region $0\le\omega\le 250\,$meV. The least squares fit
determined the parameters $I^2=0.655$ and $\omega_{SF}=82\,$meV
using full Eliashberg theory in the fitting procedure.

\begin{figure}[tp]
  \vspace*{-7mm}
  \includegraphics[width=9cm]{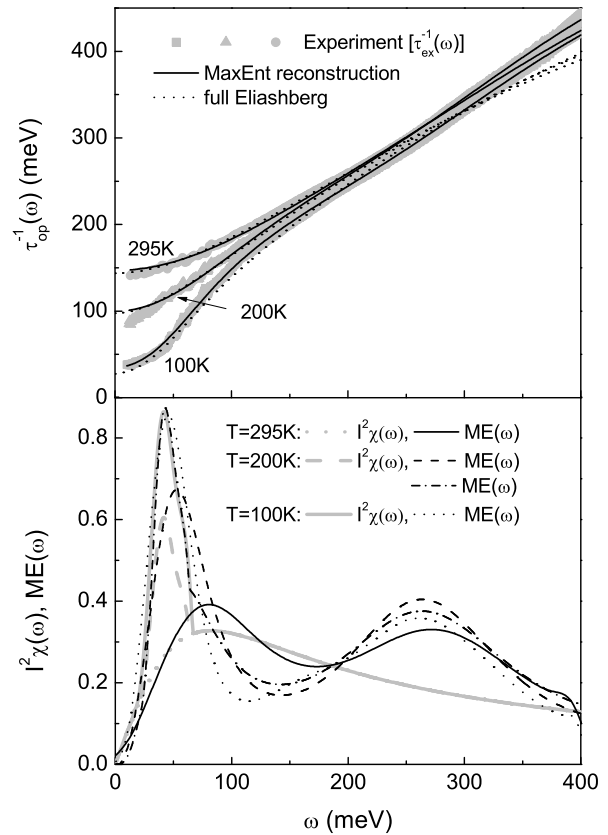}
  \caption{Top frame:
Experimental optical scattering rate data\protect{\cite{tu}}
$\tau^{-1}_{ex}(\omega)$ of Bi2212 (gray solid symbols) reconstructed
using the MaxEnt method (black solid line) for various temperatures,
namely $T=100\,$K, $200\,$K, and $295\,$K. The dotted lines
correspond to data generated by full Eliashberg theory using
the $I^2\chi(\omega)$ spectra reported by Schachinger and
Carbotte.\protect{\cite{schach1}} Bottom frame: The spectral
function ME($\omega$) (black solid line for $295\,$K, black dashed,
black dash-dotted for
$200\,$K, and black dotted for $100\,$K) as a result of the MaxEnt
inversion of the experimental data (gray solid symbols in the
top frame of this figure). The gray lines
(solid for $295\,$K, dotted for $200\,$K,
and dashed for $100\,$K) show the
$I^2\chi(\omega)$ spectra reported by Schachinger and Carbotte.%
\protect{\cite{schach1}} The black dash-dotted line presents results
of a MaxEnt inversion of the $200\,$K data (solid gray triangles in the
top frame) using the $I^2\chi(\omega)$ (gray dotted line) as the
default model.
}
  \label{fig:9}
\end{figure}

As MaxEnt turned out to be a rather powerful inversion technique
we revisit the Bi2212 data analysis. We assume the experimental data
of Tu {\it et al.} to contain uncorrelated Gaussian noise of
$\sigma=2.0$ for $T=100\,$K and $\sigma=3.0$ for $T=200\,$K and
$295\,$K. The preblur parameter was set to 5 for $T=100\,$K and to
10 for the other two temperatures. The default model was set to
0.1. The inversion is based on the application of the approximate
kernel \eqref{eq:8}. Fig.~\ref{fig:9} discusses the results of
our calculations. The top frame presents the data reconstruction
and the bottom frame the inverted spectral function ME($\omega$)
in comparison to $I^2\chi(\omega)$ spectral functions suggested by
Schachinger and Carbotte.\cite{schach1} It is quite clear that the
data reconstruction (black solid lines in the top frame of Fig.~\ref{fig:9})
is in excellent agreement with the original data (gray solid symbols)
at all temperatures. The
black dotted lines correspond to $\tau^{-1}_{op}(\omega)$ data generated
from full Eliashberg theory using the $I^2\chi(\omega)$ spectral
functions presented in the bottom frame of Fig.~\ref{fig:9} by
gray lines, namely solid for $100\,$K, dashed
for $200\,$K, and dotted for $295\,$K. The full Eliashberg results
follow the data rather nicely in the energy range $0\le\omega\le%
250\,$meV and then deviate systematically to smaller values
for energies $> 250\,$meV. Thus, the MMP form alone is not
sufficient to explain the energy dependence of $\tau^{-1}_{ex}(\omega)$
in the whole energy range $0\le\omega\le 400\,$meV.

Comparing the spectral functions ME($\omega$) to the $I^2\chi(\omega)$
spectra demonstrates rather good agreement at low energies for
$T=100\,$K (black dotted line, gray solid line). Both spectra show
a pronounced peak at $43\,$meV and they even agree in height and
width of the peak which is rather fortuitous. At $T=200\,$K
(black dashed line, gray dashed line) both
spectra develop a less pronounced peak with the peak in ME($\omega$)
shifted away from $43\,$meV to higher energies. Such a shift towards
higher energies with increasing temperatures has already been observed
in the analysis
of computer generated $\tau^{-1}_{op}(\omega)$
data for YBCO$_{6.95}$, Fig.~\ref{fig:6},
and this peak in the ME$(\omega$)
could very well correspond to a $43\,$meV peak in the `real' spectrum.
Finally, at $T=295\,$K (black solid line, gray dotted line) both spectra
agree in showing a rather flat MMP like structure peaked around
$82\,$meV with no indication of a resonance peak. This analysis
corroborates the results reported by Tu {\it et al.}\cite{tu} and
by Schachinger and Carbotte.\cite{schach1}

It is quite important to notice that all ME($\omega$) spectra develop
a second structure of comparable height around $\sim 260\,$meV for
all temperatures. Such a structure is missing in the $I^2\chi(\omega)$
spectra. This additional structure is required for a faithful
reconstruction of the experimental $\tau^{-1}_{ex}(\omega)$
data at higher energies. To include one more check we repeated
the inversion of the $T=200\,$K $\tau_{ex}^{-1}(\omega)$ data
using MaxEnt but now without preblur and with the default model
set to the $I^2\chi(\omega)$ spectrum for $200\,$K (gray dashed
line in the bottom frame of Fig.~\ref{fig:9}). The result
is presented by the black dash-dotted line in the bottom frame of
Fig.~\ref{fig:9}. In this case
the optical resonance can be found at $43\,$meV in contrast to
the calculation with the constant default model (black dashed line)
but it is now significantly enhanced.
\begin{figure}[tp]
  \vspace*{-5mm}
  \includegraphics[width=9cm]{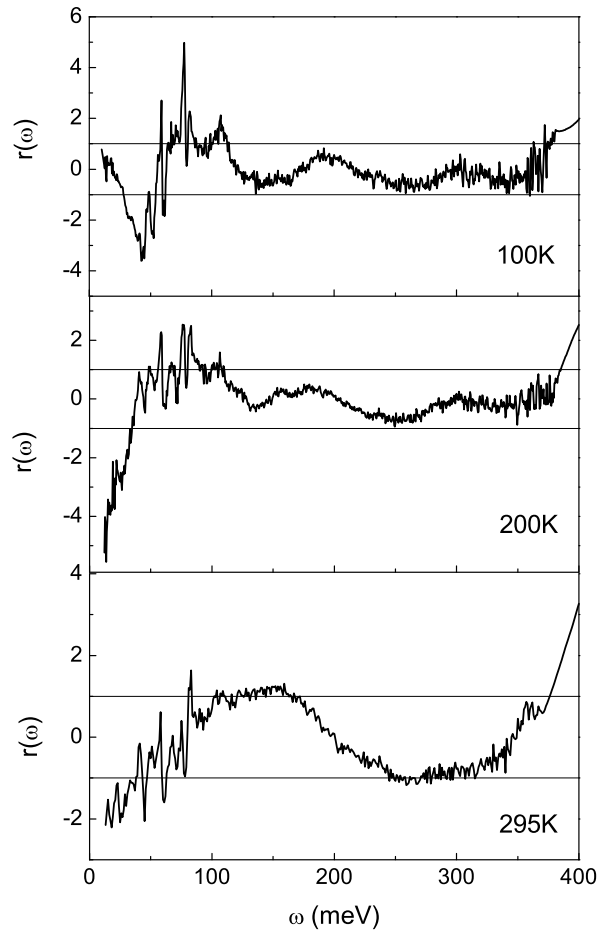}
  \caption{The residual $r(\omega)$ of the data reconstruction
presented in the top frame of Fig.~\protect{\ref{fig:9}}. The
top frame is for $T=100\,$K, the middle frame for $200\,$K,
and the bottom frame for $295\,$K.}
  \label{fig:10}
\end{figure}
Even the kink in the $I^2\chi(\omega)$ spectrum at about $60\,$meV
is reproduced in ME($\omega$). What is important, though, is the
fact that the additional high energy structure around $260\,$meV
appears again with approximately the same strength as in all other
results. Thus, it seems to be a real and new feature not
captured by a simple MMP form and this
proves that the charge carrier-exchange boson spectral function
$I^2\chi(\omega)$ in the cuprates
has non-zero contributions up to at least $400\,$meV, a
property which cannot be explained by a pure phonon mechanism.

It is interesting to note in closing that, for instance,
ME($\omega$) for $T=295\,$K can be described using a simple model,
namely the original MMP form which is replaced for $\omega>170\,$meV
by a second MMP form peaked at $\omega_{SF}=260\,$meV. A least
squares fit to $\tau^{-1}_{ex}(\omega,T=295\,$K) using full Eliashberg
theory provides an $I^2=0.55$ for this second MMP form and
agreement with the data is achieved over the whole energy range
$0\le\omega\le 400\,$meV. This is a second, independent proof
of the existence of this additional high energy structure in
$I^2\chi(\omega)$ for optimally doped Bi2212.

In Fig.~\ref{fig:10} the residual $r(\omega)$ of our
analysis of the Bi2212 data is presented. The top frame is for $100\,$K, the
middle frame for $200\,$K, and the bottom frame for $295\,$K.
The residual clearly shows a stochastic component which is much
smaller than the assumed values for $\sigma$ in the energy region
$100\le\omega\le 350\,$meV. It can be identified as a noise
contribution. There is obviously, another slowly oscillating
contribution to $r(\omega)$ which is almost identical in 
its frequency dependence
at $100\,$ and $200\,$K but it doubles its period
at $295\,$K. This contribution is very likely to be a background
signal $B(\omega)$ generated by the experimental equipment.

\begin{figure}[tp]
  \vspace*{-7mm}
  \includegraphics[width=9cm]{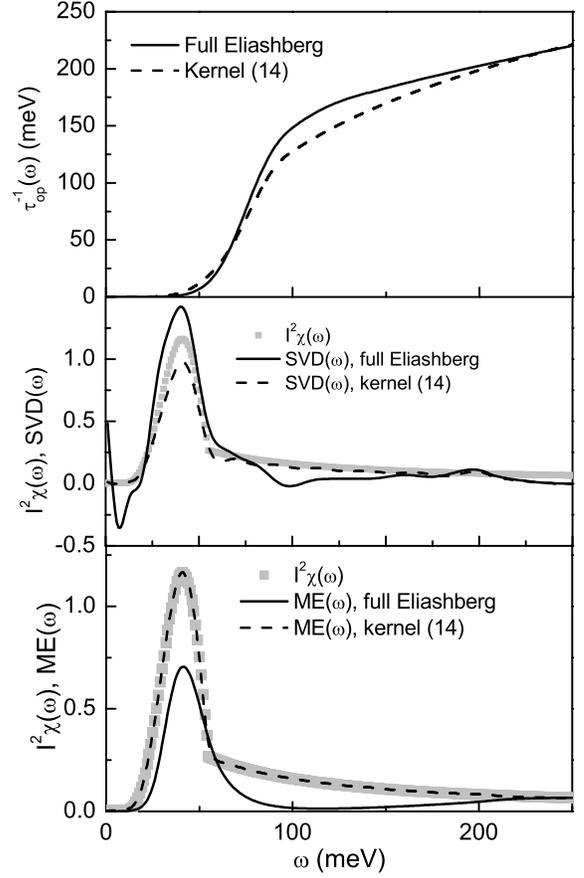}
  \caption{Top frame: The computer generated
superconducting state optical scattering
rate $\tau^{-1}_{op}(\omega)$ at $T=10\,$K calculated from full Eliashberg
theory (solid line) or from kernel \protect{\eqref{eq:11}} (dashed
line) using the $I^2\chi(\omega)$ spectrum shown as the gray solid
lines in the middle and bottom frame of this figure.
Middle frame: The gray solid line corresponds to the spectral
function $I^2\chi(\omega)$ reported by CSB for superconducting YBCO$_{6.95}$
at $T=10\,$K. The solid line shows the spectral function SVD($\omega$)
as a result of an SVD inversion of the full Eliashberg result
(solid line in the top frame of this figure). The dashed line shows
the same but now the scattering rate generated by kernel
\protect{\eqref{eq:11}} is used as input. Bottom frame: The same
as the middle frame. ME($\omega$) is the result of a MaxEnt inversion.
 }
  \label{fig:11}
\end{figure}

We proceed to investigate the inversion of superconducting state
$\tau^{-1}_{op}(\omega)$ data for YBCO$_{6.95}$. We use
for computer generated $\tau^{-1}_{op}(\omega)$ data the spectral
function $I^2\chi(\omega)$ reported by CSB for $T=10\,$K. It was derived from
experimental superconducting state $\tau^{-1}_{op}(\omega)$ data
reported by
Basov {\it et al.}\cite{bas1} at $T=10\,$K for an optimally doped,
twinned YBCO$_{6.95}$ single crystal. This $I^2\chi(\omega)$ is based
on the normal state $I^2\chi(\omega)$ for YBCO$_{6.95}$ which is an
MMP form \eqref{eq:12} with $I^2=0.83$ and $\omega_{SF} = 20\,$meV
(gray squares in Figs.~\ref{fig:6} and \ref{fig:7}).
Superimposed is a pronounced peak at $\omega=41\,$meV which was
found by applying the second order derivative method to the
experimental data. The final form, shown using gray solid squares
in the middle and bottom frame of Fig.~\ref{fig:11}, was
established by a fit of full Eliashberg
$\tau^{-1}_{op}(\omega)$ results to experiment.
This spectrum which extends to $400\,$meV was applied by
Schachinger {\it et al.}\cite{schach7} within full Eliashberg
formalism. The authors demonstrated that it was possible to
reproduce numerous superconducting state properties of
optimally doped YBCO$_{6.95}$ in their temperature and
energy dependence within experimental errors. The need for
$I^2\chi(\omega)$ to extend to several hundred meV has also been
reviewed by Basov and Timusk.\cite{bas5}

The top frame of Fig.~\ref{fig:11} presents computer generated
superconducting state $\tau^{-1}_{op}(\omega)$ data
at $T=10\,$K
as a function of energy. The solid line gives the result of a
full Eliashberg calculation using the solutions of Eqs.~\eqref{eq:Bx}
and \eqref{eq:B4} on the basis of the spectral function $I^2\chi(\omega)$
just described. Eliashberg theory also provides a value for
the zero temperature gap amplitude
$\Delta_0 = 22.03\,$meV. The dashed line presents the optical
scattering rate as calculated using kernel \eqref{eq:11}, the
above value for $\Delta_0$, and the same spectral function
$I^2\chi(\omega)$. The two results differ substantially in the
energy region $70 \le\omega\le 200\,$meV.

The results of an SVD inversion are shown in the middle frame
of Fig.~\ref{fig:11}. The solid line presents the spectral
function SVD($\omega$) found from inverting the full Eliashberg
results (solid line in the top frame of this figure) on the basis
of kernel \eqref{eq:11} using $\Delta_0 = 22.03\,$meV as an external
parameter. The dashed line corresponds to the inversion of the
scattering rate generated by kernel \eqref{eq:11} using the same
value for $\Delta_0$ (dashed line in the top frame of this figure).
In both cases the svs threshold was set to
$10^{-2}$. The agreement of both spectra SVD($\omega$) with the
original $I^2\chi(\omega)$ (gray solid line)
is rather poor keeping in mind that
the inversion is based on computer generated data.

The bottom frame of Fig.~\ref{fig:11} is organized as the middle
frame of this figure. It presents spectra ME($\omega$)
as a result of a MaxEnt inversion. For the inversion of
$\tau^{-1}_{op}(\omega)$ represented by the dashed line in the top frame of
this figure an error bar of
$\sigma = 0.01$ was assumed and no noise was added to the
data. The inversion was performed using historical MaxEnt with
$\gamma^2 = N_1$ as convergence criterion. The resulting spectral function
ME($\omega$) is represented by a dashed line. The agreement with
the spectrum $I^2\chi(\omega)$ (solid gray line) employed to generate
$\tau^{-1}_{op}(\omega)$ is perfect as was to be expected.
The inversion of full Eliashberg
theory generated $\tau^{-1}_{op}(\omega)$ data (solid line in the
top frame of this figure) based on the same $I^2\chi(\omega)$
is less successful as the resulting spectrum ME($\omega$) (solid
line) demonstrates. For the inversion an
error of $\sigma=0.7$ was attached to the $\tau^{-1}_{op}(\omega)$
data but we did not add noise. Furthermore, $\Delta_0$ had
to be reduced
to $21\,$meV in order to keep the peak in ME($\omega$) at $41\,$meV. The peak
height is now grossly underestimated and the spectrum
ME($\omega$) does no longer reproduce the normal state background spectrum.
Such a result was expected because of the pronounced differences
particularly in this energy region between the full Eliashberg
theory generated data and the data generated using kernel
\eqref{eq:11}.

We proceed and study the application of the MaxEnt inversion
based on kernel \eqref{eq:11} using experimental superconducting
state $\tau^{-1}_{op}(\omega)$ data. The top frame of
\begin{figure}[tp]
  \vspace*{-7mm}
  \includegraphics[width=9cm]{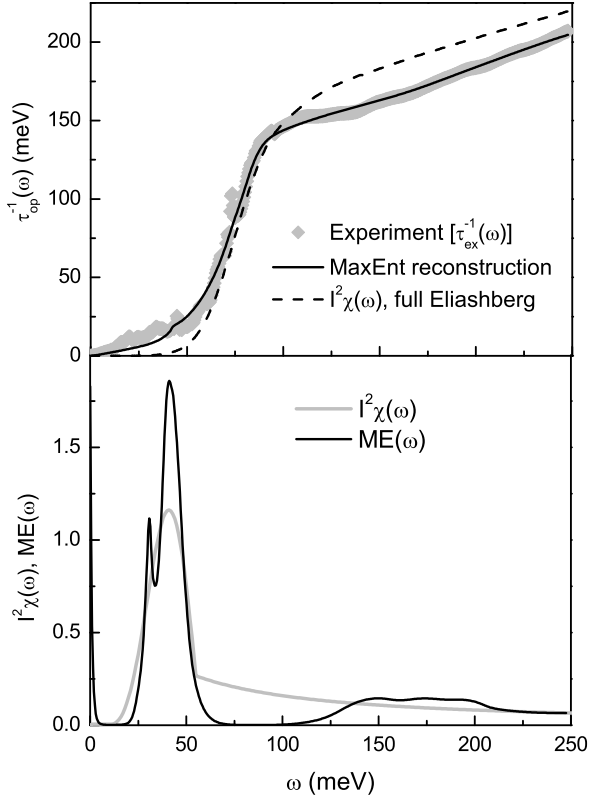}
  \caption{Top frame: The superconducting state optical scattering
rate $\tau^{-1}_{ex}(\omega)$ at $T=10\,$K for an optimally doped,
twinned YBCO$_{6.95}$ single crystal.
The solid gray diamonds present the original data by
Basov {\it et al.}\protect{\cite{bas1}}. The solid line shows
the MaxEnt reconstruction of the input data while the dashed line
shows the result of a full Eliashberg theory calculation based
on the spectrum $I^2\chi(\omega)$ reported by CSB for $T=10\,$K.
Bottom frame: The gray solid line presents the $I^2\chi(\omega)$
spectrum suggested by CSB for superconducting YBCO$_{6.95}$ at $T=10\,$K.
The solid line shows the spectrum ME($\omega$) as a result of the
inversion of the experimental data by Basov {\it et al.}%
\protect{\cite{bas1}} represented by
solid gray diamonds in the top frame of this figure.
}
  \label{fig:12}
\end{figure}
of Fig.~\ref{fig:12} presents
the original data by Basov {\it et al.}\cite{bas1} reported for an
optimally doped, twinned YBCO$_{6.95}$
single crystal at $T=10\,$K (solid gray diamonds).
The inversion of this data is performed using historical MaxEnt
based on kernel \eqref{eq:11} with $\gamma^2=N_1$ as criterion of convergence.
An error bar of $\sigma=3.5$ was attached to the data and the default
model was set to 0.05. The resulting spectrum
ME($\omega)$ is presented as black solid line in the bottom frame of
Fig.~\ref{fig:12}. This spectrum was found using $\Delta_0=21\,$meV.
This allowed to place the main peak in ME($\omega$) at
$41\,$meV. It is obvious that the inverted spectrum
ME($\omega$) differs substantially from the
$I^2\chi(\omega)$ spectrum (gray solid line) reported by CSB.
It shows peak splitting of the resonance peak at
$41\,$meV and the low energy
peak ($\omega< 5\,$meV) in ME($\omega$) is caused by an attempt
to use MaxEnt to
extrapolate to very small energies which are not supported by the
$\tau^{-1}_{ex}(\omega)$ data. Nevertheless, the data reconstruction
(black solid line in
the upper frame of Fig.~\ref{fig:12}) is excellent. We added for
comparison (dashed line in the upper frame of Fig.~\ref{fig:12})
the optical scattering rate as generated by full Eliashberg theory
using $I^2\chi(\omega)$.
The agreement with the data does not seem to be good enough to
justify the particular shape of $I^2\chi(\omega)$ discussed above.
Nevertheless, one has to keep in mind that all calculations
presented here are performed without including impurities,
i.e.: in the pure case limit. Adding impurities improves the agreement
between full Eliashberg theory and experiment substantially.\cite{schach1}

As a last example we present the reconstruction of
experimental superconducting
state $\tau^{-1}_{ex}(\omega)$ data reported by Tu {\it et al.}\cite{tu}
for an optimally doped Bi2212 single crystal at $T=6\,$K.
\begin{figure}[tp]
  \vspace*{-7mm}
  \includegraphics[width=9cm]{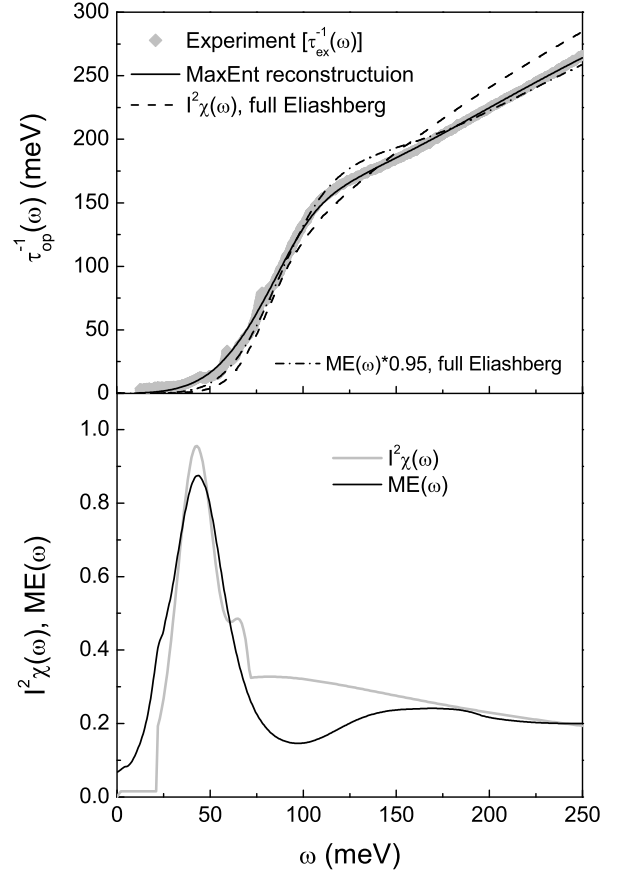}
  \caption{The same as Fig.~\protect{\ref{fig:12}} but now for
a superconducting optimally doped Bi2212 single crystal at
$T=6\,$K. The data have been reported by Tu {\it et al.}%
\protect{\cite{tu}} The additional dash-dotted line in
the top frame describes the result of a full Eliashberg calculation
based on the ME($\omega$) spectrum which was down scaled by a factor
0.95.
  \label{fig:13}
}
\end{figure}
The results are presented in Fig.~\ref{fig:13} which is organized
the same way
as Fig.~\ref{fig:12}. For the MaxEnt data reconstruction an
error bar determined by $\sigma=3.0$ was attached to the data.
Historical MaxEnt with $\gamma^2=N_1$ as criterion for convergence
was applied. The default model was set to 0.2.
The inversion is based on kernel \eqref{eq:11} and the
spectrum $I^2\chi(\omega)$ reported by Schachinger and Carbotte%
\cite{schach1} for $T=6\,$K is shown as a gray solid line
in the bottom frame of Fig.~\ref{fig:13} for comparison.
It contains a peak
at $43\,$meV and an MMP form \eqref{eq:12} as background with
$I^2=0.655$ and $\omega_{SF} = 82\,$meV. In this case the agreement
between the inverted spectrum ME($\omega$) (solid line in the
bottom frame of Fig.~\ref{fig:13}) and $I^2\chi(\omega)$ is much
better in comparison to YBCO$_{6.95}$. This confirms the analysis of
Schachinger and Carbotte\cite{schach1} as well as a previous
analysis of Bi2212 data by Schachinger and Carbotte\cite{schach2}
based on data published by Puchkov {\it et al.}\cite {puchkov}
We added one more result to the top frame of Fig.~\ref{fig:13}
presented by a dash-dotted line. It corresponds to the result of
a full Eliashberg calculation based on the ME($\omega$) spectrum
(down scaled by a factor of 0.95) instead of $I^2\chi(\omega)$.
The agreement with experiment is now very good. In particular,
the `overshoot' right after the main rise in the optical
scattering rate\cite{bas5} is better resolved than in the
original Eliashberg calculation (dashed line). Thus, the existence
of this overshoot is an indication that a dip exists in the
$I^2\chi(\omega)$ spectrum immediately following the resonance
peak. This dip will, of course, not be as pronounced as it
appears in ME($\omega$) because the major part of it stems from
the differences between the approximate kernel \eqref{eq:11}
and full Eliashberg theory.
If the ME($\omega$) spectrum is, furthermore,
employed to calculate the
zero temperature gap amplitude $\Delta_0$ within full Eliashberg
theory a value of $32\,$meV is found, in excellent
agreement with experimental results.\cite{zas1,zas2}
We also note that the optical scattering rate
$\tau^{-1}_{ex}(\omega)$ in the top frame of Fig.~\ref{fig:12}
(gray solid diamonds) develops a moderate overshoot following the
main rise around $80\le\omega\le110\,$meV. Thus, also in this
case not all of the dip which follows the resonance peak in the
ME($\omega$) spectrum can be attributed to the differences in the
approximate kernel \eqref{eq:11} and full Eliashberg theory.

\subsection{The least squares fit method}
\label{ssec:3.3}

It has been pointed out in the previous subsection that the
least squares fit method has already been applied rather
successfully to invert $I^2\chi(\omega)$ spectra from experiment
using full Eliashberg theory together with additional information gathered
by other means. This method is rather clumsy to handle and time
\begin{figure}[tp]
  \vspace*{-7mm}
  \includegraphics[width=9cm]{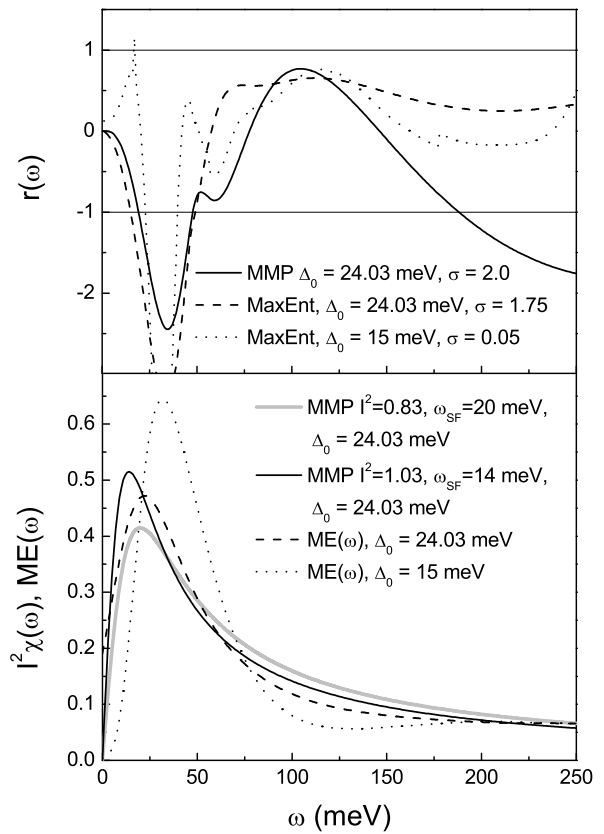}
  \caption{Top frame: The residual $r(\omega)$.
The solid line corresponds to a least square
fit of an MMP form to computer generated optical scattering
rate data generated by full Eliashberg theory, $\Delta_0=24.03\,$meV
and $\sigma=2.0$ was assumed. The dashed line corresponds to a MaxEnt analysis
of the same data for $\sigma=1.75$ keeping $\Delta_0$ fixed and
the dotted line is for  $\sigma=0.05$ and $\Delta_0=15\,$meV. The main
part of the various $r(\omega)$ curves is within $\pm1$ indicating
perfect data reconstruction within the assumed error. Bottom
frame: The spectrum $I^2\chi(\omega)$ (gray solid
line) is the input spectrum for the full Eliashberg calculation.
The solid line gives the inverted
spectrum as a result of a least squares fit to the data, the
dashed line the result of a MaxEnt inversion both with
$\Delta_0 = 24.03\,$meV. Finally, the
dotted line shows the spectrum ME($\omega$) which resulted from
an optimal data reproduction using MaxEnt but now $\Delta_0=15\,$meV.
}
  \label{fig:14}
\end{figure}
consuming as one cannot develop a closed algorithm which allows
one to fit parameters directly given some standard
deviation $\sigma$ which plays the role of the nuisance parameter.
Therefore, we want to study the least squares fit method
based on the approximate kernel \eqref{eq:11} for the
superconducting state of a $d$-wave superconductor. The
$\tau^{-1}_{op}(\omega)$ data are generated from full Eliashberg
theory for the superconducting state at $T=10\,$K using for
$I^2\chi(\omega)$ an MMP form \eqref{eq:12} with $I^2=0.83$
and $\omega_{SF} = 20\,$meV (gray solid line in the bottom
frame of Fig.~\ref{fig:14}).
The zero temperature
gap $\Delta_0 = 24.03\,$meV. The least squares fit method
is now applied to determine $I^2$ and $\omega_{SF}$ of an
MMP form by a least squares fit to $\tau^{-1}_{op}(\omega)$
in the energy region $0\le\omega\le 250\,$meV. The error bar
attached to the input data is given by $\sigma=2.0$. (This
particular value of the standard deviation appears to be a
realistic value for the reconstruction of experimental data as was
demonstrated in the previous subsection.)
No noise was added. A consistent data
reconstruction was achieved with the parameters $I^2=1.03$ and
$\omega_{SF}=14\,$meV. This becomes apparent from Fig.~\ref{fig:14}
in which the results of the least squares method are illustrated.
The solid line in the top frame of this figure shows the residual
$r(\omega)$, Eq.~\eqref{eq:residual}, which is on average well
within the assumed $\sigma$. This insures a correct data reconstruction.
The bottom frame of this figure compares the least squares fit
spectrum (solid line) to the original $I^2\chi(\omega)$ indicated
by a gray solid line. It is interesting to compare the areas
under these two spectra, they are $41.5\,$meV and $42\,$meV,
respectively, a difference of about 1\%. The parameter $\lambda$
which is two times the first inverse moment of $I^2\chi(\omega)$
is also a good parameter to compare. We get $\lambda=2.48$ and
$3.09$, respectively.

Figure.~\ref{fig:14} contains additional information. We use the
MaxEnt method to generate an `educated guess' for a later least
squares fit to data based on full Eliashberg theory. If we use
$\Delta_0=24.03\,$meV and assume $\sigma=1.75$ historical
MaxEnt reproduces the input data almost equally well as our
least squares fit. (Dashed line in the top frame of Fig.~\ref{fig:14}.)
The resulting spectrum ME($\omega$) (dashed line in the bottom
frame of Fig.~\ref{fig:14}) has its peak at a slightly higher
energy ($\sim 24\,$meV) as compared to the original $I^2\chi(\omega)$
but otherwise, the input spectrum is reproduced rather well,
albeit not by an MMP form. The area under this spectrum is $40\,$meV
and $\lambda=3.22$, again close to the result of the least squares
fit `inversion'. We also include, for comparison, the result of a
MaxEnt deconvolution with the emphasis on optimal data reconstruction.
We reduce the error bar on the input data to $\sigma=0.05$ and
use $\Delta_0$ as a parameter to be adjusted in order to achieve
this goal. An almost perfect reproduction is possible if
$\Delta_0$ is reduced to $15\,$meV. This becomes apparent from
the residual $r(\omega)$ shown as a dotted line in the top frame
of Fig.~\ref{fig:14}. The resulting spectrum ME($\omega$) is
presented by a dotted line in the bottom frame.
The peak is now shifted to much higher energies, it is wider,
and is greater in height as compared to the original $I^2\chi(\omega)$.
The area under the spectrum is $40.2\,$meV and $\lambda=1.92$.
This demonstrates clearly the approximate nature of
kernel \eqref{eq:11} and that a value for $\Delta_0$ found from
optimum data reproduction is not physically significant. Instead,
one has to treat $\Delta_0$ as an external parameter to the
inversion process which is to be
determined by other means like, for instance, scanning
tunneling microscopy.\cite{zas1,zas2}

While the various methods of inversion described in this section
lead to significant differences in the value of $\lambda$ obtained,
the area under the spectral density varies very little. The differences
in $\lambda$ can be traced mainly to variations in the position
of the main fluctuation peak and, therefore, it is recommended
that independent information on the position of this peak be
used, for example from neutron scattering.

As a result of this study one can say that given additional information,
like the value of the zero temperature gap amplitude,
both methods, least squares
fit and MaxEnt, result in comparable spectra,
nevertheless, they differ qualitatively and
quantitatively from the `real' spectrum $I^2\chi(\omega)$. This
emphasizes the role of additional information beyond the optical
data for a successful data analysis.

\section{Conclusion}
\label{sec:4}

There exists a well established formalism that relates the
electron-phonon spectral density $\alpha^2F(\omega)$ to the
infrared conductivity. It applies to the superconducting as well as
normal state and involves the Eliashberg equations plus a Kubo
formula which gives the optical conductivity
$\sigma_{op}(\omega)$ from Green's functions.
While such a formalism is not as well justified in the case of
other boson exchange mechanisms such as spin fluctuations,
it has, nevertheless, been useful to apply it as a first
approximation with appropriate essential modifications such as
$d$-wave gap symmetry. The resulting equations are, however, rather
complicated and simplified, lowest order perturbation theory
expressions for the
relationship between spectral density
$I^2\chi(\omega)$ and optical conductivity have played an
important role particularly if a main aim is to extract
qualitative rather than quantitative information on the size
and main features of the $I^2\chi(\omega)$ for a given set of
optical data. If, however, accurate quantitative information
is desired a full Eliashberg formulation cannot be avoided.
In this paper we provided comparison between numerical results
for the optical scattering rate $\tau^{-1}_{op}(\omega)$ based
on the exact equations and results generated from several often
used approximate relations between conductivity and spectral
density including a recent generalization which applies to a
superconductor at $T=0$ with $d$-wave symmetry.

Another important issue discussed in detail is the accuracy,
advantages, and limitations of various numerical methods which are
needed to invert data even within the limitations of approximate
formulas for the optical scattering rate. These equations relate
the optical scattering rate measured in infrared experiments to
the desired spectral density $I^2\chi(\omega)$ through an
convolution integral involving a known, specified kernel
$K(\Omega,\omega;T)$ multiplied by $I^2\chi(\omega)$.
A second derivative technique of the optical scattering rate, often
used, is also considered.

In the normal state at low temperatures the second derivative
technique applied to computer generated optical scattering rates
based on Eliashberg theory reproduces well the spectral
function $\alpha^2F(\omega)$ of Pb. There is a slight overestimate
of the longitudinal
and transverse phonon peaks as well as a small shift to higher
energy. The region between the peaks is slightly underestimated and
unphysical tails occur beyond the Debye energy but these are to
be ignored. If the same data is inverted using either SVD or MaxEnt,
the agreement between the spectral function of Pb and
and the deconvoluted spectral functions remains good
even though both inversion methods are based on an approximate lowest
order perturbation theory expression for the relation between
spectral function and scattering rate while the
$\tau^{-1}_{op}(\omega)$ data is
based on Eliashberg relations. It is noted, however, that
MaxEnt does somewhat better than SVD which introduces additional
oscillations into the resulting spectral density not present
in the $\alpha^2F(\omega)$ of Pb.
As the temperature is increased in the normal state all three
methods begin to fail and at $T=50\,$K the two peak structure of
the Pb $\alpha^2F(\omega)$ can no longer be resolved.

For the
cuprates in their normal state an often used spectral function
$I^2\chi(\omega)$ is the MMP form of the NAFFL model. This
function is characterized by a spin fluctuation frequency
$\omega_{SF}$ and a coupling constant between spin susceptibility
and the charge carriers. The MMP form has a peak at
$\omega=\omega_{SF}$ but is rather smooth and extends to energies
of order $100\,$meV. For such a relatively unstructured spectrum,
even at $100\,$K, the deconvoluted spectral function is much
closer to the original MMP form than was the case for Pb.
For the SVD inversion, however,
there is a spurious splitting of the peak at $\omega=\omega_{SF}$
(for small $\omega_{SF}$) and some additional oscillations
occur. These oscillations
are also seen in the case of the MaxEnt inversion but they are
less prominent and quite small.
Of the two inversion methods considered above
MaxEnt is to be preferred because it has the advantage that few
assumptions, namely a default model and error bars attached to
the data, will result in a deconvoluted spectrum which allows
excellent data reconstruction. This spectrum can also be expected
to contain, at least qualitatively, all the main features of
the `real' spectrum.

As a result of all this, inversion of normal state
$\tau^{-1}_{ex}(\omega)$ data, even at high temperatures, can
lead to useful quantitative results if the boson exchange
spectral function is expected to be rather smooth with little
structure. The application of methods based on the
approximate kernel \eqref{eq:8} is to be favored because
there are only little differences between this kernel and
full Eliashberg theory.

For the low temperature superconducting state use of the
Eliashberg equations for the calculation of the optical
scattering rate $\tau^{-1}_{op}(\omega)$ along with numerical
inversion based on approximate lowest order perturbation theory
simplified formulas leads to larger quantitative differences
between the $\alpha^2F(\omega)$ spectrum of lead and
deconvoluted spectrum than were noted
for the normal state of lead. Nevertheless, both SVD and MaxEnt
methods yield very useful, qualitative and even semi
quantitative information on the shape and size of $\alpha^2F(\omega)$
with the MaxEnt method, again, to be preferred.

Inversion of experimental data on the optical scattering rate
(as opposed to computer generated data) already exists in the
literature in a few cases and these are based on full Eliashberg
solutions and the Kubo formula. These inversions, however,
proceed through a least squares fit of the scattering rates
assuming a specific mathematical form for $I^2\chi(\omega)$
characterized by a few parameters, namely an MMP form with a low
frequency cutoff plus a resonance peak at a specified frequency.
Such fits have had considerable success when applied to normal
and superconducting state in optimally doped cuprates. Attempts
have been made since additional complications arise due to the
emergence of the pseudogap. As yet no consensus exists as to the
origin of this pseudogap and, thus, modeling its effect remains
controversial.

Even though, as noted above, the SVD and MaxEnt methods are limited
due to inaccuracies in the approximate second order perturbation
form used to relate optical scattering rate to spectral density
(instead of the complete Eliashberg analysis), we have used MaxEnt
inversions to confirm the previously obtained least squares fits.
All qualitative features of the resulting $I^2\chi(\omega)$
function, namely, coupling to an optical resonance at low energy
and to a background extending even beyond $400\,$meV are
confirmed and new features have been unveiled. This
demonstrates the usefulness of such techniques which make
no a priory assumption on the shape or size of the underlying spectral
density. To get best results the value of the zero temperature
gap amplitude $\Delta_0$ (obtained in some other way) should
be used rather than varied arbitrarily to get a best fit.
The numerical spectra for $I^2\chi(\omega)$ could be
used as a first step in a more complicated inversion process
which would involve further refinements along the lines of
the least squares fit procedure described above. Other constraints
such as known properties of the superconducting state could be
added as well in the fit. It is not clear,
however, that this is necessary and worthwhile for the oxides
where the mechanism is not the electron-phonon interaction
and additional complications such as a breakdown of the Migdal
theorem may arise.

\section*{Acknowledgment}
 
Research supported in part by the Natural Sciences and Engineering
Research Council of Canada (NSERC), by the Canadian
Institute for Advanced Research (CIAR), and by Fonds zur
F\"{o}rderung der Wissenschaftlichen Forschung (Vienna) under contract
No. P15834-PHY. We thank D.N.~Basov for interest and discussion.
Two of us (D.N. and E.S.) are indebted to
W. von der Linden for discussions, support, and his interest
in this work.

\appendix
\section{The Maximum Entropy Method for Data Analysis}
\label{app:A}

The direct inversion of Eq.~\eqref{eq:5} constitutes an ill-posed problem.
Therefore, there are many different `solutions' (varying orders of magnitude)
that fit the data within the error bars. The most general solution to this
problem is the calculation of the {\em posterior} probability distribution
(pdf) $p({\bf a}\vert{\bf t},{\cal I})$ of possible
solutions {\bf a} given the data {\bf t} and all additionally available
background information ${\cal I}$, i.e.: the matrix
\mbox{\boldmath $K$} which is defined by the underlying theoretical
model, the background function $B(\omega)$, the noise contribution
$\eta(\omega)$, etc. Bayesian probability theory \cite{sivia}
provides the consistent framework for such a fully probabilistic
description. Bayes' theorem provides the relation
\begin{equation}
  p({\bf a}\vert{\bf t},{\cal I}) = \frac{p({\bf t}\vert{\bf a},
  {\cal I})p({\bf a}\vert{\cal I})}{p({\bf t}\vert{\cal I})},
  \label{eq:A1}
\end{equation}
which relates the posterior $p({\bf a}\vert{\bf t},{\cal I})$ to the
{\em likelihood} pdf $p({\bf t}\vert{\bf a},{\cal I})$ and the
{\em prior} pdf $p({\bf a}\vert{\cal I})$. Finally, the
denominator $p({\bf t}\vert{\cal I})$ ensures proper normalization
of the posterior.
The likelihood comprises the model definition
and the error statistics of the data. Its knowledge is an essential
prerequisite for any data analysis.
The prior, on the other hand, should incorporate all available
information of the problem at hand. In the particular case discussed
here, the only known constraint is the positivity of the
function values $a_j \equiv \alpha^2 F(\Omega_j)$.

Skilling\cite{skilling} showed that the most uninformative prior in
this case is the Maximum Entropy prior:
\begin{equation}
  \label{eq:A2}
  p({\bf a}\vert\alpha,{\cal I}) = \exp(\alpha S)
  \left(\prod\limits_{j=1}^{N_2}\sqrt{a_j}\right)^{-1}. 
\end{equation}
Here, $\alpha$ is a renormalization (nuisance) parameter and $S$ is
the generalized Shannon-Jaynes entropy\cite{sivia}
\begin{equation}
  \label{eq:A3}
  S = \sum\limits_{j=1}^{N_2} \left[a_j - m_j - a_j \log \frac{a_j}{m_j}
  \right].
\end{equation}
It measures the distance of the candidate vector {\bf a} from the so-called
{\em default model} vector ${\bf m} = \{m_j\vert j=1,\ldots,N_2\}$,
which represents the most probable solution prior the
observation of any data. In case of insufficient background information it
should be chosen constant, i.e.: $m_j \equiv \text{const},\,\forall j$.
Nevertheless, it is adamant
to check its influence on the solution, as certain
features of the solution might not be supported by the data but instead just
reflect the initial assumption of the default model.

The regularization parameter $\alpha$ determines the relative
influence of the prior compared to the likelihood. In the limit
$\alpha\to\infty$ one obtains ${\bf a}\to{\bf m}$ as the most probable
solution;
for $\alpha\to 0$, on the other hand, one gets the maximum likelihood
solution which will be meaningless for ill-conditioned problems.
Within conventional approaches, regularization parameters such as $\alpha$
are often fixed by hand. Apart from ad-hoc settings, a sensible choice
is to adjust the regularization parameter such that the expectation
value of the misfit $\gamma^2$ is reproduced.\cite{footnote}
In case of an $N$-dimensional uncorrelated normal
distribution the misfit is described by the $\chi^2$-distribution with
$N$ degrees of freedom and has mean $\langle\gamma^2\rangle = N$ and
variance $\text{var}(\gamma^2) = 2N$. Historically, the criterion
$\gamma^2=N$ was employed first in order to fix the parameter
$\alpha$ (historical MaxEnt). However, one has to keep in mind that
the solution might change dramatically if the regularization parameter
$\alpha$ is tuned such that $\gamma^2$ varies between
$N-\sqrt{2N}\le\gamma^2\le N+\sqrt{2N}$.

In principle, the regularization parameter $\alpha$ can be determined
consistently within Bayesian probability theory by computing the
most probable value $\alpha$ which maximizes the probability
$p(\alpha\vert{\bf t},{\cal I})$ given the data {\bf t}. (This is the
classical MaxEnt of Ref.~\onlinecite{gull}.) Unfortunately, the
calculation of $p(\alpha\vert{\bf t},{\cal I})$ involves high
dimensional integrals which can only be evaluated using rather crude
simplifications. The approximation usually applied\cite{gull}
tends to overfit the data as $p(\alpha\vert{\bf t},{\cal I})$ is
systematically overestimated for small $\alpha$ which results in a
too small $\hat{\alpha}$-value at which $p(\alpha\vert{\bf t},{\cal I})$
has its maximum as a function of $\alpha$.
Von der Linden\cite{wvl} suggested a different approximation scheme
that partly corrects these deficiencies and yields results similar to the
historic criterion.

For some data sets analyzed in the study, we found that all methods
to determine the value of $\alpha$ suffered from oscillations
(`ringing') due to overfitting. This has been observed for other
applications as well.\cite{fischer1,fischer2} To a certain extent,
this ringing is intrinsic to the MaxEnt prior which explicitly
treats all points of the reconstruction {\bf a} as uncorrelated.

In order to enforce smoothness of the solution Skilling\cite{skilling2}
suggested the introduction of a `hidden image' {\bf h} which is
blurred by a Gaussian
\begin{equation}
  \label{eq:preblur}
  a_j = \sum\limits_k B_{jk}h_k,\qquad
  B_{jk} = \frac{1}{\sqrt{2\pi b^2}}\exp\left[-\frac{(x_j-x_k)^2}{2b^2}
  \right].
\end{equation}
Here, the $x_j$ designate the abscissas of $a_j$ and $h_j$. The vector
{\bf a} enters the likelihood, while {\bf h} is used to compute the
entropy $S$. The blur-width $b$ is an additional hyper-parameter that
can be determined simultaneously with $\alpha$ by locating the
maximum of $p(b,\alpha\vert{\bf t},{\cal I})$ given the data
{\bf t} in the spirit of Ref.~\onlinecite{skilling2}.

Various choices of the blur-width $b$ can be regarded as distinct
models which have a different number of degrees of freedom (similar
to fit functions involving different numbers of parameters). For
$b\to 0$ all positive discrete representations {\bf a} can be realized
as ${\bf a}\to{\bf h}$, while in the limit $b\to\infty$ only constant
functions $a_i\equiv \text{const.}$ can be represented, i.e.: the model
has only one effective degree of freedom.

The optimal blur-width $b$ is determined by the interplay of
the likelihood  and Occam's razor\cite{sivia,fischer2} which penalizes
the complexity of the model employed and is implicit in the
calculation of $p(b,\alpha\vert{\bf t},{\cal I})$. The `penalty factor'
is the ratio of the width of the likelihood and the prior distributions.
Thus, a simpler model may be more favorable because a larger fraction
of the parameter space is likely to be realized according to data
although a more complex model fits the data better.

Unless stated otherwise, we have determined the optimal blur-width
$b$ for the MaxEnt reconstructions presented in Sec.~\ref{sec:3} as
outlined above. For the computation of $p(b\vert{\bf t},{\cal I})$
we chose a flat prior $p(b\vert{\cal I})$ on the interval
$b_{min}\le b\le b_{max}$ with $b_{min}\sim x_2-x_1$ and
$b_{max}\sim x_N-x_1$.

The MaxEnt method obviously allows for an explicit treatment of
ambiguous solutions and it allows prior knowledge to be taken
into account consistently by introducing a suitable prior pdf.
A direct inversion, like the SVD method, which may be badly
conditioned or may involve uncontrolled approximations, is
avoided. Finally, it is possible to obtain error estimates.
Nevertheless, it has to be pointed out that `fuzzy' constraints
such as smoothness of the output of the inversion process make
a definition of the prior pdf rather complicated.\cite{drose}

\section{Eliashberg equations}
\label{app:B}

\begin{widetext}
The generalized, clean limit Eliashberg Equations which play
an important role in this study are
\begin{subequations}
\label{eq:Bx}
\begin{eqnarray}
  \label{eq:B1}
    \tilde{\Delta}(\nu+i0^+;\theta) &=& \pi Tg
  \sum\limits_{m=0}^\infty\cos(2\theta)\left[\lambda(\nu-i\omega_m)+
  \lambda(\nu+i\omega_m)\right]h(i\omega_m)\\
 &&+i\pi g\int\limits^\infty_{-\infty}\!dz\,\cos(2\theta)
  I^2\chi(z)\left[n(z)+f(z-\nu)\right]h(i\omega_m\to\nu-z+i0^+),\nonumber
\end{eqnarray}
and, in the renormalization channel,
\begin{eqnarray}
  \tilde{\omega}(\nu+i0^+) &=& \nu+i\pi T%
  \sum\limits_{m=0}^\infty\left[\lambda(\nu-i\omega_m)-
  \lambda(\nu+i\omega_m)\right]g(i\omega_m)\nonumber\\
  &&+i\pi\int\limits^\infty_{-\infty}\!dz\,
   I^2\chi(z)\left[n(z)+f(z-\nu)\right]g(i\omega_m\to\nu-z+i0^+).
  \label{eq:B2}
\end{eqnarray}
\end{subequations}
Here
\[
 h(i\omega_m) = \left\langle
 \frac{\tilde{\Delta}(i\omega_m;\theta)\cos(2\theta)}
  {\sqrt{\tilde{\omega}^2(i\omega_m)+\tilde{\Delta}^2(i\omega_m;
  \theta)}}\right\rangle_\theta,\quad
 g(i\omega_m) = \left\langle
    \frac{\tilde{\omega}(i\omega_m)}
  {\sqrt{\tilde{\omega}^2(i\omega_m)+\tilde{\Delta}^2(i\omega_m;
  \theta)}}\right\rangle_\theta,
\]
and the parameter $g$ allows for a possible difference in spectral
density between $\tilde{\omega}$ and $\tilde{\Delta}$ channels.
It is fixed to get the measured value of the critical temperature.
In the above $\tilde{\Delta}(i\omega_m;\theta)$ is the pairing
energy evaluated at the fermionic Matsubara frequencies
$\omega_m = \pi T(2m-1), m = 0,\,\pm1,\,\pm2,\ldots$; $f(z)$
and $n(z)$ are the Fermi and Bose distribution, respectively. The
renormalized Matsubara frequencies are $\tilde{\omega}(i\omega_m)$.
The analytic continuation to real frequencies $\nu$ of the above is
$\tilde{\Delta}(\nu+i0^+;\theta)$ and $\tilde{\omega}(\omega%
+i0^+)$. The
brackets $\langle\cdots\rangle_\theta$ are the angular average
over $\theta$, and
  $\lambda(\nu) = \int^\infty_{-\infty}\!d\Omega\,
   \alpha^2F(\Omega)/(\nu-\Omega+i0^+)$.  
Eqs.~\eqref{eq:Bx} are a set of nonlinear coupled equations for the
renormalized pairing potential $\tilde{\Delta}(\nu+i0^+;\theta)$
and the normalized frequencies $\tilde{\omega}(\nu+i0^+)$ with
the gap
  $\Delta(\nu+i0^+;\theta) = \tilde{\Delta}(\nu+i0^+;\theta)/
  Z(\nu)$,
where the renormalization function $Z(\nu)$ was introduced in the
usual way as $\tilde{\omega}(\nu+i0^+) = \nu Z(\nu)$. To get
the $s$-wave version of these equations $g$ is set equal to one and all
$\cos(2\theta)$ factors are to be omitted with no average over
the polar angle $\theta$. A Coulomb pseudopotential $\mu^\ast$ must also
be introduced in Eq.~\eqref{eq:B1}.

The optical conductivity follows from knowledge of $\tilde{\omega}$
and $\tilde{\Delta}$. The formula to be evaluated is
\begin{equation}
  \sigma_{op}(T,\nu) = \frac{\Omega^2_p}{4\pi}\frac{i}{\nu}
     \left\langle
      \int\limits_0^\infty\!d\omega\,\tanh\left(\frac{\beta\omega}{2}
        \right)\left[
        J(\omega,\nu)- J(-\omega,\nu)
      \right]\right\rangle_\theta.  \label{eq:B4}
\end{equation}
The function $J(\omega,\nu)$ is given by
\begin{eqnarray}
  \label{eq:B6}
  2J(\omega,\nu) &=& \frac{1-N(\omega;\theta)N(\omega+\nu;\theta)-
   P(\omega;\theta)P(\omega+\nu;\theta)}{E(\omega;\theta)+
   E(\omega+\nu;\theta)}\nonumber\\
   &&+\frac{1+N^\ast(\omega;\theta)N(\omega+\nu;\theta)+
   P^\ast(\omega;\theta)P(\omega+\nu;\theta)}{E^\ast(\omega;\theta)-
   E(\omega+\nu;\theta)},
\end{eqnarray}
with
  $E(\omega;\theta) = \sqrt{\tilde{\omega}^2
   (\omega+i0^+)-\tilde{\Delta}^2
   (\omega+i0^+;\theta)}$,
  $N(\omega;\theta) = \tilde{\omega}(\omega+i0^+)/
   E(\omega;\theta)$, and
  $P(\omega;\theta) = \tilde{\Delta}(\omega+i0^+;\theta)/
   E(\omega;\theta)$.
Finally, the star refers to the complex conjugate.
\end{widetext}


\begin{thebibliography}{99}
\bibitem{mars3}F.\ Marsiglio and J.P.\ Carbotte, in {\it The Physics of
Superconductivity: Conventional and High-Tc Superconductors},
edited by K.-H.~Bennemann and J.B.~Ketterson (Springer, Berlin, 2003)
Vol. I, p. 233.
\bibitem{carb1} J.P. Carbotte, Rev. Mod. Phys. \textbf{62}, 1027 (1990).
\bibitem{allen} P.B. Allen, \prb \textbf{3}, 305 (1971).
\bibitem{carb4}J.P.~Carbotte, E.~Schachinger, and J.~Hwang,
\prb \textbf{71}, 054506 (2005).
\bibitem{mcmillan} W.L. McMillan and J.M. Rowell, \prl
\textbf{14}, 68 (1965).
\bibitem{farnworth} B. Farnworth and T. Timusk, \prb \textbf{10},
2799 (1974); \prb \textbf{14}, 5119 (1976).
\bibitem{tomlinson} P. Tomlinson and J.P. Carbotte, \prb \textbf{13},
4738 (1976).
\bibitem{boza1} B. Mitrovi\'{c} and M.A. Fiorucci, \prb
\textbf{31}, 2694 (1985).
\bibitem{boza2}B.~Mitrovi\'{c} and S.~Perkowitz, \prb \textbf{30},
6749 (1984).
\bibitem{mars1} F. Marsiglio, T. Startseva, and J. P. Carbotte,
Physics Lett. A \textbf{245}, 172 (1998).
\bibitem{pines1} A. V. Chubukov, D. Pines, and J. Schmalian, in:
{\it The Physics
of Superconductivity: Conventional and High-T$_c$ Superconductors},
edited by K.-H. Bennemann and J. B. Ketterson (Springer, Berlin, 2003),
Vol. I, p. 495.
\bibitem{pines2} P. Monthoux and D. Pines, Phys. Rev. B \textbf{47},
6069 (1993); Phys. Rev. B \textbf{49}, 4261 (1994).
  \bibitem{kulic1}M.L.\ Kuli\'c, Phys.\ Rep. {\bf 338}, 1 (2000).
  \bibitem{zeyher}R.~Zeyher and M.L.~Kuli\'c, \prb {\bf 53},
    2850 (1996); {\bf 54}, 8985 (1996).
  \bibitem{kulic2}M.L.~Kuli\'c and R.~Zeyher, \prb {\bf 49},
    R4395 (1994); Physica C {\bf 199-200}, 358 (1994);
    {\bf 235-240}, 358 (1994).
  \bibitem{weger}M.~Weger, B.~Barbelini, and M.~Peter,
    Z.\ Phys.\ B {\bf 94}, 387 (1994).
  \bibitem{weger1}M.~Weger, M.~Peter, and L.P.~Pitaevskii,
    Z.\ Phys. B {\bf 101}, 573 (1996).
  \bibitem{kulic3}O.V.\ Danylenko, O.V.\ Dolgov, M.L.\ Kuli\'c, and
     V.\ Oudovenko, Euro.\ Phys.\ Jour.\ B-Cond.\ Matter {\bf 9},
     201 (1999).
  \bibitem{kulic4}M.L.~Kuli\'c and O.V.~Dolgov, in {\it
   High Temperature Superconductivity}, edited by S.~Barnes,
   J.~Ashkenazi, J.~Cohn, and F.~Zuo, AIP Conf. Proc. No. 483
   (AIP, Woodbury, NY, 1999), p. 63.
\bibitem{schach5}E.~Schachinger and J.P.~Carbotte, 
in: {\it Models and Methods of High-TC Superconductivity: some Frontal
Aspects}, edited by J.K.~Srivastava and S.M.~Rao,
(Nova Science, Hauppauge, NY, 2003),  Vol. II, pp. 73.
\bibitem{zas1}J.F.~Zasadzinski, L.~Ozyuzer, N.~Miyakawa, K.E.~Gray,
D.G.~Hinks, and C.~Kendziora, \prl \textbf{87}, 067005 (2001).
\bibitem{zas2}J.F.~Zasadzinski, L.~Coffey, P.~Romano, and
Z.~Yusof, \prb \textbf{68}, 180504(R) (2003).
\bibitem{zas3}J.F.~Zasadzinski, L.~Ozyuzer, L.~Coffey, K.E.~Gray,
D.G.~Hinks, and C.~Kendziora, cond-mat/0510057 (unpublished).
\bibitem{kam1}A.~Kaminski, M.~Randeria, J.C.~Campuzano, M.R.~Norman,
H.~Fretwell, J.~Mesot, T.~Sato, T.~Takahashi, and K.~Kadowaki,
\prl \textbf{86}, 1070 (2001).
\bibitem{campu1}J.C.~Campuzano, H.~Ding, M.R.~Norman, H.M.~Fretwell,
M.~Randeria, A.~Kaminski, J.~Mesot, T.~Takeuchi, T.~Sato, T.~Yokoya,
T.~Takahashi, T.~Mochiku, K.~Kadowaki, P.~Guptasarma, D.G.~Hinks,
Z.~Konstantinovic, Z.Z.~Li, and H.~Raffy, \prl \textbf{83}, 3709 (1999).
\bibitem{norm}M.R.~Norman, M.~Eschrig, A.~Kaminski, and
J.C.~Campuzano, \prb \textbf{64}, 184508 (2001).
\bibitem{eschrig}M.~Eschrig and M.R.~Norman, \prl \textbf{85}, 3261 (2000).
\bibitem{cuk}T.~Cuk, F.~Baumberger, D.H.~Lu, N.~Ingle, X.J.~Zhou,
H.~Eisaki, N.~Kaneko, Z.~Hussain, T.P.~Devereaux, N.~Nagaosa,
and Z.X.~Shen, \prl textbf{93}, 117003 (2004).
\bibitem{carb3} J. P. Carbotte, E. Schachinger, and D. Basov,
Nature (London) \textbf{401}, 354 (1999).
\bibitem{tu}J.J.~Tu, C.C.~Homes, G.D.~Gu, D.N.~Basov, and
M.~Strongin, \prb \textbf{66}, 144514 (2002).
\bibitem{schach3}E.~Schachinger, J.J.~Tu, and J.P.~Carbotte, \prb
\textbf{67}, 214508 (2003); Physica C \textbf{364-365}, 13 (2001).
\bibitem{hwang}J.~Hwang, T.~Timusk, and G.D.~Gu, Nature (London)
\textbf{427}, 714 (2004).
\bibitem{mmp} A. J. Millis, H. Monien, and D. Pines, Phys. Rev. B
\textbf{42}, 167 (1990).
\bibitem{schach2}E.~Schachinger and J.P.~Carbotte, \prb \textbf{62},
9054 (2000).
\bibitem{schach1}E.~Schachinger and J.P.~Carbotte, J.~Phys.~Stud.
\textbf{7}, 209 (2003).
\bibitem{dorde} S.V. Dordevic, C.C. Homes, J.J. Tu, T. Valla,
M. Strongin, P.D. Johnson, G.D. Gu, and D.N. Basov, \prb
\textbf{71}, 104529 (2005).
\bibitem{shulga1} S.V. Shulga, O.V. Dolgov, and E.G. Maksimov,
Physica C \textbf{178}, 266 (1991).
\bibitem{sharap} S.G. Sharapov and J.P. Carbotte, \prb \textbf{72},
134506 (2005).
\bibitem{carb2} J.P. Carbotte and E. Schachinger, Ann. Phys. (in print).
\bibitem{hwang1}J.~Hwang, J.~Yang, T.~Timusk, S.G.~Sharapov,
J.P.~Carbotte, D.A.~Bonn, R.~Liang, and W.N.~Hardy, cond-mat/0505302
(unpublished).
\bibitem{sivia}See for instance: D.S.~Sivia, {\it Data Analysis},
Clarendon Press, Oxford (1996).
\bibitem{nash} J. C. Nash, {\it Compact Numerical Methods for
Computers: Linear Algebra and Function Minimalization,} Adam Hilger,
Bristol (1990), p. 30.
\bibitem{jaynes}E.T. Jaynes, Phys. Rev. \textbf{106}, 620 (1957);
\textbf{108}, 171 (1957).
\bibitem{joyce} R.R. Joyce and P.L. Richards, \prl
\textbf{24}, 1007 (1970).
\bibitem{branch} D. Branch and J. P. Carbotte, Can. J. Phys. \textbf{77},
531 (1999); Jour. Supercond. \textbf{12}, 667 (1999).
\bibitem{bas1}D.N.~Basov, A.V.~Puchkov, R.A.~Hughes, T.~Strach,
J.~Preston, T.~Timusk, D.A.~Bonn, R.~Liang, and W.N.~Hardy, \prb
\textbf{49}, 12165 (1994).
\bibitem{schach7}E.~Schachinger, J.P.~Carbotte, and D.N.~Basov,
Europhys. Lett. \textbf{54}, 380 (2001).
\bibitem{bas5} D.N.~Basov and T.~Timusk, \rmp \textbf{77}, 721 (2005).
\bibitem{puchkov}A.~Puchkov, D.N.~Basov, and T.~Timusk,
J.~Phys.: Condens. Matter \textbf{8}, 10\ 049 (1996).
\bibitem{skilling}J. Skilling, in: {\it Maximum Entropy and Bayesian Methods},
edited by J. Skilling (Kluwer, Dortrecht, 1989), pp. 45.
\bibitem{footnote}We would like to point out that
in the MaxEnt literature the symbol $\chi^2$ is used instead of $\gamma^2$.
\bibitem{gull}S.F.~Gull, in: {\it Maximum Entropy and Bayesian Methods},
edited by J. Skilling (Kluwer, Dortrecht, 1989), pp. 53.
\bibitem{wvl}W.~von~der~Linden, R.~Preuss, and V.~Dose, in:
{\it Maximum Entropy and Bayesian Methods},
edited by W.~von~der~Linden, V.~Dose, R.~Fischer, and R.~Preuss
(Kluwer, Dortrecht, 1989), pp. 285.
\bibitem{fischer1}R.~Fischer, Anal. Bioanal. Chem. \textbf{374}, 619 (2002).
\bibitem{fischer2}R.~Fischer, W.~von~der~Linden, and V.~Dose,
in: {\it MAXENT'96 - Proceedings of the Maximum Entropy Conference 1996},
edited by M.~Sears, V.~Nedeljkovoc, N.E.~Pendock, and S.~Sibisi,
(NMB Printers, Port Elisabeth, 1996), p. 21 and references therein.
\bibitem{skilling2}J.~Skilling, in: {\it Maximum Entropy in Action},
edited by B.~Buck and V.A.~Macauly (Calrendon Press, London, 1991), p.19.
\bibitem{drose}V.~Dose, Rep. Prog. Phys. \textbf{66}, 1421 (2003).
\end{thebibliography}
\end{document}